\documentclass[a4paper,natbib,12pt]{article}
\pdfoutput=1
\usepackage{amssymb}
\usepackage{amsmath}
\usepackage{epsfig}
\usepackage{graphicx}
\usepackage{color,ulem}
\usepackage{hyperref}
\usepackage{multirow}
\usepackage{ulem}

\makeatletter

\@addtoreset{equation}{section}
\def\be{\begin{equation}}
\def\ee{\end{equation}}
\def\bea{\begin{eqnarray}}
\def\eea{\end{eqnarray}}
\def\eq{\begin{eqnarray}}
\def\eqx{\end{eqnarray}}

\def\({\left(}
\def\){\right)}
\def\<{\left<}
\def\>{\right>}

\def\be{\begin{equation}}
\def\ee{\end{equation}}
\def\ben{\begin{eqnarray}}
\def\een{\end{eqnarray}}
\def\({\left(}
\def\){\right)}
\def\<{\left<}
\def\>{\right>}
\def\!{\right|}
\def\|{\left|}

\def\[{\left[}
\def\]{\right]}

\def\+{\bar}

\def\A{{\cal{A}}}

\def\bx{{\bf{x}}}

\usepackage{hyperref}
\hypersetup{colorlinks,%
  citecolor=blue,%
  linkcolor=blue,%
 }

\begin{document}

\begin{titlepage}
\vskip1cm
\begin{flushright}
% UOSTP {\tt 1512001}
\end{flushright}
\vskip0.25cm
\centerline{
\bf %\large 
Analyzing Planar Galactic Halo Distributions with Fuzzy/Cold Dark Matter Models 
}
\vskip1cm \centerline{ \textsc{
 Sangnam Park,$^{\, \tt a \dagger}$ Dongsu Bak,$^{\, \tt a,b\dagger}$   Jae-Weon Lee,$^{\, \tt c}$ Inkyu Park$^{\, \tt a,b}$ } }
\vspace{1cm}
 \centerline{\sl a) Natural Science Research Institute,
University of Seoul, %Siripdaero 163,
Seoul 02504,  Korea}
 \vskip0.4cm
\centerline{\sl  b) Physics Department,
University of Seoul, Seoul 02504, Korea}
 \vskip0.3cm
 \centerline{\sl c) Department of Electrical and Electronic Engineering, Jungwon University}
 \centerline{85 Munmuro, Goesan, Chungbuk 28024, Korea}
   \vskip0.4cm
 \centerline{
%E-mail: 
{\tt \small (u98parksn@gmail.com,\,dsbak@uos.ac.kr,\,scikid@jwu.ac.kr,\,icpark@uos.ac.kr)}
}

\let\thefootnote\relax\footnote{$^\dagger$ These authors contributed equally to this work.}
\addtocounter{footnote}{-1}
\vskip1.7cm
%\centerline{\today}
%\vspace{1.75cm}
\centerline{ABSTRACT} \vspace{0.75cm}
{
\noindent
We perform a numerical comparison between the fuzzy dark matter model
and the cold dark matter model, focusing on 
formation of satellite galaxy planes around massive galaxies.
Such galactic dynamics with controlled initial subhalo configurations are 
investigated
using GADGET2 for the cold dark matter and PyUltraLight for the fuzzy dark matter,
respectively. 
We demonstrate that satellite galaxies in the fuzzy dark matter side have a tendency
to form more flattened and corotating satellite systems than in the cold dark matter side mainly due to the dissipation by
the gravitational cooling effect of the fuzzy dark matter.
Our simulations
with the fuzzy dark matter typically show
the minor-to-major axis ratio $c/a$ 
of the satellite galaxy planes to be $0.21 \sim  0.30$; This  well matches the current observed value for the Milky Way.
}
\end{titlepage}
%%%%%%%%%%%%%%%%%%%%%%
%\maketitle

%%%%%%%%%%%%%%%%%%%%%%%%%
\section{Introduction}\label{sec1}
%Dark matter (DM) is one of the main ingredients of the universe providing a gravitational attraction to form cosmic structures~\cite{Silk:2016srn}.
%The most popular DM model is

 Numerical
simulations based on the cold dark matter (CDM) model successfully reproduce  the observed large-scale structures
 of the universe. However, it
encounters difficulties in explaining  small-scale structures around  galactic scales. For example, the CDM model
predicts a cusped central halo density which is not observed
and too many dwarf galaxies around massive host galaxies than the observed
\cite{Salucci:2002nc,Navarro:1995iw,deblok-2002,crisis}.

On the other hand, %the plane of
the satellite-plane problem is another serious 
 small-scale challenge for the CDM model (see  \cite{Pawlowski:2018sys,Pawlowski:2021ipt} for a review).
Observations indicate that
 a substantial fraction of satellite %dwarf 
 galaxies around the Milky Way, Andromeda and Centaurus A
  co-orbit in flattened planar structures \cite{Muller,2020MNRAS.499.3755S}.
  At least 10 highly flattened planes of dwarf galaxies have been discovered
  \cite{2019MNRAS.490.3786L}.
 This is in sharp contrast with cosmological simulations with the CDM, which show
  much more random distributions and motions of satellite galaxies.
{  The probability of finding such flattened $and$ co-rotating satellite systems with the CDM is very low \cite{Pawlowski:2018sys,Muller} although the statistic is  controversial.
%
% -- ${\cal O}(0.1\%)$ 
}

It is not easy to explain these satellite planes theoretically in the CDM model.
{  Disks in astronomical objects such as a protoplanetary disk or a black hole accretion disk form when falling matter such as gas loses its energy by a dissipation (cooling) mechanism while keeping the angular momentum of the matter. 
For a given angular momentum, the lowest energy state is a disk orthogonal to the rotation axis \cite{disk}.
Similarly, to form a flat disk-like structure small galactic halos should
 lose their gravitational energy by some dissipation  mechanism
when they  fall towards  heavy central galaxies.
 However, the typical collisionless CDM model lacks an efficient mechanism for this to happen. Due to the large scales $({\cal O}(10^2 {\rm kpc)})$  baryonic processes could not strongly affect the satellite planes.}

As an alternative to the CDM, there is a growing interest in the fuzzy dark matter (FDM) 
\cite{Fuzzy,1983PhLB..122..221B, 1989PhRvA..39.4207M,Sin:1992bg,jwlee1},
also known as ultra-light axion or scalar field dark matter.
In this model,  dark matter (DM) particles satisfying the Schr\"odinger-Poisson equation
have an ultra-light mass $m\simeq  10^{-22} {\rm \, eV/c^2}$ and
a long de Broglie wavelength $\lambda=2\pi \hbar/mv\simeq  {\rm \,kpc}$, where $v$ is the velocity of the dark matter
particles.
 (See \cite{Hui:2016ltb,Lee:2017qve} for more references.)
 { Unlike CDM particles, condensed FDM particles behave as a coherent wave, and
 due to the quantum uncertainty principle the FDM has a typical length scale $\lambda$ which suppresses the formation of DM structures smaller than a galaxy core  
 and helps resolve the aforementioned small-scale issues \cite{Hui:2016ltb}.}

 While the CDM model has no efficient dissipation mechanism to make the observed   satellite systems,  a gravitational cooling effect in the FDM model
  provides a unique and efficient dissipation mechanism for them.
  The gravitational cooling \cite{Guzman:2006yc, Bak:2018pda,Bak:2020det} is a mechanism for relaxation
by ejecting part of the 
FDM density and carrying out excessive kinetic energy and  momentum.
In this mechanism
colliding FDM halos can have interfering wave profiles which
 contain high momentum modes. These modes can easily escape the gravitational
potential of the halos carrying out energy.

 The main aim of this paper is to show that halos in the FDM model have a tendency to form a more flattened and corotating satellite plane than in the standard collision-less CDM model\footnote{Mixed state solutions of the
Schr\"odinger-Poisson equation were suggested to be a cause of the anisotropic concentration
of satellite galaxies~\cite{Solis-Lopez:2019lvz}, which is different from our proposal.}.
 To show this tendency we perform numerical simulations of formation of toy galactic systems
 using Gadget2 for the CDM and  PyUltraLight for the FDM, respectively.
 Since it is challenging to simulate a full-fledged 
 cosmological %cosmic scale 
 structure formation with the FDM
 due to the wave nature of the model, we restrict ourselves to studying the differences in structure formation at galactic scales with controlled
 initial conditions for seed satellite %dwarf 
 galaxies in the two models. 
(There are cosmological scale hydrodynamical simulations 
with FDM and baryon matter  for structure formation \cite{Mocz:2019pyf}.)
 %as observed.

% Our approach in this paper will lead to %proposes
% another solution to the satellite plane problem.

In  Section~{\ref{sec2}} we describe our numerical simulation setup %methods 
for formation of galactic systems in the FDM  and the CDM models. 
We shall apply the numerical results to find the motion of satellite galaxies using the so called Mulguisin halo finder.
%The Sec.~{\ref{sec4}}
% contains
Section~{\ref{sec4}} contains the results of our numerical study. We analyze and
 compare resulting satellite distributions of the two DM models.
In  Section~{\ref{sec5}} we compare our results with the observed astronomical data.
The last section is devoted to  concluding remarks.
%discussions.

%
%
%
\section{Simulation Setup}\label{sec2}
%\subsection{Simulation of Dark Matter Halos}\label{sec2.1}
%
%
%

In this section, we would like to describe our simulation setup for the above two DM models. 
As was mentioned already, we are not trying to do a full-fledged cosmological simulation, which is certainly beyond scope of our current study. Rather, we setup specific initial configurations %(put in a large box) 
for the late-time formation of galactic satellite systems, focusing on the purely gravitational dynamics in a fixed flat background. 
Our primary aim is to see the resulting differences between the FDM and the CDM dynamics.

\subsection{  FDM subhalo seeds  }\label{sec2.1}

%The dynamics of our FDM halo model is described by the Schr\"odinger-Newton equation
Our FDM dynamics are governed by the Schr\"odinger-Poisson equations
\begin{align}\label{SNeq}
i\hbar \partial_t \psi(\bx,t ) & = -\frac{\hbar^2}{2m}\nabla^2 \psi(\bx,t )  +m V(\bx,t ) \psi(\bx,t )  %\nonumber 
\\
 \nabla^2 V(\bx,t )&=4\pi GM_{\rm tot} \, |\psi|^2(\bx,t ) 
\end{align} 
where $m\, (=10^{-22} {\rm  eV/c^2})$ is the mass of ultralight scalar particles,  $G$ for the Newton constant, and $M_{\rm tot}$ denotes the total mass of the galactic system of interest. 
%\sout{The wave function is normalized as $\int d^3 \bx\,|\psi|^2 =1$.} 
{ The wave function $\psi(\bx, t)$ is a complex function of spacetime normalized as $\int d^3 \bx\,|\psi|^2 
=1$, %and 
the corresponding FDM mass density  then given by $\rho= M_{\rm tot}|\psi|^2$,  and $V$ for the gravitational potential produced by the density $\rho$. The $m$ value is the fiducial mass
suggested to solve the small scale problems
 such as the missing satellite problem. If we use a  much larger value for $m$
we expect the wave nature of FDM becomes less relevant and
the difference of FDM from CDM becomes less prominent.}

As an initial subhalo configuration, we shall use the ground state solution (see e.g. \cite{Hui:2016ltb}) of the Schr\"odinger-Poisson system with a  mass $M_s$, which will be  called  a `soliton' below  for simplicity \cite{chavanis}.
 Let us denote this soliton wave function centered at ${\bf x}=0$ by
$h({\bf x};M)$, whose explicit functional form is known numerically\footnote{We take this to be real without introducing any extra phase at this stage.  In  \cite{Schive:2014hza},  an approximate fit function of this soliton configuration was introduced as 
$h_{\rm fit}=0.51411/r^{3/2}_{1/2} /(1+ 0.19191 r^2/r^2_{1/2})^4$
 where $r_{1/2}$ is the half mass radius. This form will not be used in our simulation below.}. 
%An initial phase may be assigned to each soliton in general %but in the following we shall set it to zero.
It is also known that the half mass radius $r_{1/2}$ of the soliton is given by
%\sout{$r_{1/2}\simeq 3.9251\,\l_c\M/M_s$.}
$r_{1/2}\simeq 3.3541\times 10^8\, {\rm kpc} \, M_\odot  /M_s$ \cite{Hui:2016ltb}.  % \sim 1.068$ kpc. 
 This soliton configuration is spherically symmetric, stable under a small perturbation, and may work as a seed for late-time satellite galaxies in our simulation. Although we are not explicitly presenting it in this note, a moving soliton 
solution with a constant velocity  is rather well known  in terms of the above wave function $h({\bf x};M_s)$ \cite{Edwards:2018ccc}. Using this, one may obtain an initial wave function \cite{Edwards:2018ccc}
\be\label{solwave}
\psi_{sol} ({\bf x};{\bf x}_0; {\bf v}_0;M_s)=h({\bf x}-{\bf x}_0;M_s)\, e^{i\frac{ m }{\hbar}{\bf v}_0 \cdot ({\bf x}-{\bf x}_0)+i\varphi_0 }
\ee
describing an initial condition for a soliton with an initial velocity ${\bf v}_0$ centered at an initial 
position ${\bf x}_0$. The phase $\varphi_0 (\in [0,2\pi)\,)$ will be assigned randomly for each soliton below. In this note, we shall take the initial subhalo mass to be  $M_s %= 25 \times \M
 \simeq 3.142 \times 10^8 M_\odot$ leading to %, as a initial state of a subhalo in our FDM halo model. The soliton of this mass have 
%the half mass radius 
$r_{1/2}%$ = $3.9251\,\l_c\M/M 
\simeq 1.068$ kpc. { This choice, which is slightly larger than the typical satellite mass of
nearby galaxies, was made because, through interactions with the rest, 
%including the centroid, 
the  seeds generically  lose their mass in the formation of their satellite system. }

For our FDM simulation, we shall use the python  pseudo-spectral solver, PyUltraLight, developed in \cite{Edwards:2018ccc}, which is publicly available. In this package, the whole system is 
placed in a box of size $L$ with a periodic boundary condition $x_i \sim x_i +L$.   
We shall take the box size to be $L %=30 \ell_c 
\simeq 204.0\,\,$kpc and the number of the grid points ($N_g$) in each direction is set to $900$.
Hence our spatial resolution becomes $\Delta x = L/N_g \simeq 0.2267$ kpc.
%}
%For our choice of optimized time step, see \cite{Bak:2020det}. 
We choose the simulation 
%use our choice of  
time-step, %\Inkyu{
$\Delta t %= {\tau_c}/{3000}
\simeq 0.7861$ Myr, %},  
which is slightly smaller than the default value of the python code, 
%\sout{$\Delta t_{\rm d} = \frac{\Delta x %^2}{\pi \ell_c^2} \tau_c
%\simeq {\tau_c}/{2827}$.} 
{ $\Delta t_{\rm d} 
\simeq 0.8342$ Myr.}
%with $\Delta x = L_{box}/N_g$ 2827.4.
See \cite{Bak:2020det} for a detailed explanation of a similar choice. 
{ In PyUltraLight, the full initial wave function is prepared by a simple superposition of the initial soliton wave functions describing initial  subhalo seeds with given velocities. To avoid any significant overlap contributions, we separate initial positions such that one may  ignore their overlap contributions. With this initial configuration, the Schr\"odinger-Poisson equations  are solved with the above specified boundary condition, which produces time-series sets of wave-function maps leading to an evaluation of relevant physical quantities such as density maps.}

\subsection{CDM subhalo seeds
}\label{sec2.2}

We now turn to our simulation setup in the CDM side. 
Here we shall basically perform a gravitational $N$-body simulation using Gadget2 \cite{Springel:2005mi} which is also publicly available and we would like to imitate our initial  FDM halo configurations as much as possible.   
Specifically, we set the mass of the initial subhalo to be $M_s$ as before and use the same sets of initial positions and velocities as the FDM counterparts. However, a stable solitonic configuration is   not available in the CDM side. %unfortunately. 
We instead prepare an initial CDM subhalo based on the NFW profile \cite{Navarro:1995iw}
whose mass density function is given by
\begin{align}
 \rho_{NFW}(r) &= \frac{\rho_0}{\frac{r}{R_s} (1 + \frac{r}{R_s})^2}
\end{align}
where $R_s$ is the length scale of the profile.
In particular we generate $N_p$ ($=2000$) CDM particles with a mass $M_s/2000$, whose 
initial positions are randomly allocated  following the NFW profile with a cut-off scale 
% taken to be 
set by $3 R_s$.
The two parameters, $\rho_0$ and $R_s$ are fixed by the conditions
\begin{align}
 M_s \ &=4\pi  \int_0^{3R_s} dr r^2  \rho_{NFW}(r),\\
 M_s/2 &=4\pi \int_0^{r'_{1/2}} dr r^2   \rho_{NFW}(r), 
\end{align}
where $r'_{1/2}$ will be further adjusted %appropriately 
in the following way.
We set %the initial velocities of 
these $2000$ particles to move only in angular directions,  
selected randomly, and  their speed 
to $(1-\epsilon_{h})\sqrt{{GM(r)}/{r}}$ with
a judicious choice of $\epsilon_{h}=1/30$ where $M(r)$ denotes the total %particle 
mass of particles within 
a radius $r$. Now performing a separate $N$-body simulation,  %with such initial conditions, 
we let the subhalo system evolve for  $1.0$ Gyr %$1.0\, \tau_c$ 
such that the subhalo particle system is relaxed into a dynamical equilibrium. We adjust  the initial parameter $r'_{1/2}$ such that the resulting half mass radius agrees with that of the FDM soliton ($r_{1/2}$). In fact we found 
that $r'_{1/2} \approx r_{1/2}$ with the above choice of $\epsilon_h$. We then use the resulting 
particle positions and velocities as our actual initial subhalo configuration. { This relaxed configuration slightly differs from the original NFW profile we began with. However we view that this does not matter due to the following two reasons; First the NFW profile would not be that precise configuration that follows from the CDM simulations \cite{Wyithe:2000si,Dutton:2014xda}. Second we tried other models such as the Plummer and the above soliton and found that the resulting differences are rather negligible.}

{ Gadget2 is a cosmological $N$-body simulation code based on the so-called TreeSPH code, by which one may compute the nonlinear regime of gravitational dynamics and hydrodynamics. In the simulation, with $N$ particles, we first assign their initial positions and velocities, and evaluate their gravitational time evolution by computing the gravitational force on each particle using a TreePM algorithm
(see \cite{Springel:2005mi} for the details).}

In   our % this 
CDM simulation, we again put the $N$-body system inside a box with a size $L_{box}$
together with the spatially periodic boundary condition. In the simulation with Gadget2, we also set the particle mesh grid (PMG) scale to  $L_{box}/128\simeq 1.594$ kpc. This choice 
%for our CDM halo simulation 
may offer %be considered to give 
a computational efficiency without much loss of the accuracy in the calculation %computation 
of forces based on the purely Tree algorithm. 
%{\color{red} We set the softening length (SL) scale to %be approximately  
%the value 
%$\epsilon_{2b} %\approx %2r_{200}/N_{200}$ $
%\simeq 4r_{1/2}/N_p$, which ensures preventing large-angle deflections in two-body encounters \cite{Power:2002sw}. 
%for the usually required prevention of the large angle deflections during two-body encounters. 
%We have also tested  less restrictive choices  
%SL$'$ $> \epsilon_{acc} \simeq 2r_{1/2}/\sqrt{N_p}$. %$ $(\approx r_{200}/\sqrt{N_{200}})$, 
%$This requirement is needed to prevent any strong discreteness effects where
%$\epsilon_{acc}$ plays a role of setting the upper limit of acceleration in two-body encounters
%by the mean-field value of a subhalo %\cite{Power:2002sw}. We have tried various such choices and found that 
%there are no significant differences in their performances.}
{ We set the softening length (SL) scale to value
$\epsilon_{acc} \simeq 2r_{1/2}/\sqrt{N_p}$. %$ $(\approx r_{200}/\sqrt{N_{200}})$, 
This requirement (SL $> \epsilon_{acc}$) is needed to prevent any strong discreteness effects where
$\epsilon_{acc}$ plays a role of setting the upper limit of acceleration in two-body encounters
by the mean-field value of a subhalo \cite{Power:2002sw}.
We have also tested  more smaller scale SL$'$
$ = \epsilon_{2b} %\approx %2r_{200}/N_{200}$ $
\simeq 4r_{1/2}/N_p$, which ensures preventing large-angle deflections in two-body encounters \cite{Power:2002sw}. 
%for the usually required prevention of the large angle deflections during two-body encounters. 
 We have tried various such choices and found that 
there are no significant differences in their performances.}

We make our resulting datasets of CDM density fields with the same resolution as those of
the FDM side for an unbiased comparison of the two sides. In this analysis, we use the 
Pynbody  package \cite{pynbody} which is an open source code especially suitable for 
astrophysical $N$-body  and smoothed-particle hydrodynamics 
(SPH) problems. 
We down-size the datasets of  three dimensional density fields from original 900 cells along each side of the simulation box to 450 cells, which is mainly for our computational convenience % manipulation 
while keeping the accuracy needed for an identification of the satellite galaxies.
These satellites,  in the present case,
as small clusters of DM distribution will  be identified  with Mulguisin halo finder, whose details are described   
%\sout{in the following section.}
 below.

%We made the density field data of our CDM halo model with the same resolution as that of the FDM side for unbiased comparison using the Pynbody open source package  \cite{pynbody} for astrophysical N-body and smooth particle hydrodynamic simulations. We down-sized the three dimensional density field data from 900 cells along each side of simulation box to 450 cells for convenient manipulation without significant loss of the accuracy for the identification of the satellite galaxies which are to be defined as the objects clustered with Mulguisin halo finder described in detail at the following section.

%\subsection{Measures of Degrees of Flattening}\label{sec2.2}
%The spatial distribution of the satellites in our FDM halo model is clearly different with that of satellites in the CDM one, and also the number of satellites in the simulations for our FDM halo model is about 25 percent of that for CDM model (see figure \ref{2x2DFmap} and \ref{mgsDFmap}). For these differences, comparing the degree of flattening between our CDM and FDM halo models with only a specific measure might be insufficient to conclude.  
%We describe here three frequently used measures in the studies of satellites plane, to show that halos in  our FDM halo model has a tendency to form more flattened satellite planes than in the CDM one.

\subsection{Initial halo configuration}\label{sec2.1.1}

{ In order to prepare our initial configuration
for seed subhalos, we use the Plummer model \cite{Plummer:1911zza}
whose density 
profile is described by
\begin{align}
 \rho_{P}(r) = \frac{3 M_{tot}}{4 \pi r_0^3\,  (1 + \frac{r^2}{r_0^2})^{\frac{5}{2}}}
\end{align}
where $r_0$ denotes the half light radius. This %simple 
model is chosen to simply  
%for simplicity, which may  also 
provide enough randomness for our initial halo configuration. We shall set $r_0 \simeq 10.423$ kpc, which seems an appropriate choice compared to the size of our simulation box.
Including the velocity dependence,
the distribution function becomes
\begin{align}
f_P(\bx,{\bf v}) = \frac{24 \sqrt{2}\, r_0^2}{7 \pi^3  \, G^5  M^5_{tot}}  (-E(\bx,{\bf v}))^{\frac{7}{2}}
\end{align}
if $E(\bx,{\bf v}) <0$ and $f_P(\bx,{\bf v})=0$ otherwise, where $E(\bx,{\bf v})=\frac{1}{2} v^2\negthinspace -\negthinspace
%+ V_P(r)$ with$V_P(r)= 
%-{GM_{tot}}/{\sqrt{r^2+r_0^2}}
\text{\footnotesize ${GM_{tot}}/\negthinspace{\sqrt{r^2+r_0^2}}$}
\,$.  Our initial halo configuration  will be consisting of $200$ seed subhalos.
In each set of this initial configuration, we begin by randomly generating %$200$ %seed 
subhalo positions and velocities following the above distribution function. In order to prepare an 
out-of-equilibrium configuration, we shall  then multiply an overall factor $q_1\, (<1)$ to the above generated velocity
$\vec{v}_P$ for each seed subhalo.
 For the smaller $q_1$ (and also for $q_2$ introduced right below), the initial halo configuration becomes the more tightly bound gravitationally and, of course, the more subhalos are then infalling to the central region.

Now in order to provide an overall angular momentum for our initial halo configuration, we proceed as follows. With the above prepared initial set for total 200  subhalos, we randomly select $Z$ subhalos  and add  an angular velocity $q_2\,\vec{v}_+ = q_2\, v_{rot}(r)\sin \theta \, \hat{\phi}$ $(q_2 \le 1)$ to each of them where
$v_{rot}(r)$ is the rotation velocity $\sqrt{\frac{G M_{tot}(r)}{r}}$ with $M_{tot}(r)$ denoting the sum of masses within the radius $r$ from the center of the box. To the remaining $(200-Z)$ subhalos, we add  $q_2\,\vec{v}_- =-\frac{q_2}{2} \, v_{rot}(r)\sin \theta \, \hat{\phi}$. Thus
the resulting initial velocity  becomes
\begin{align}
\vec{v}_{ini}= q_1\, \vec{v}_P + q_2\, \vec{v}_\pm
\end{align}
for the $Z$ or the remaining $(200-Z)$ subhalos, respectively. Finally %Additionally, 
we  translate, rotate and boost the above halo system %adjust uniformly the positions  and the velocities of all the subhalos 
such that  it %the resulting halo system  
has a vanishing center of mass position and velocity involving only a $z$ component of total angular momentum.
The parameters $q_2$ and $Z$ are introduced to obtain an appropriate spin of our halo. Of course, the random choice of the $Z$ subhalos may provide an extra randomness for our initial subhalo configuration.  In order to consider a well-bound system, one has to restrict the parameter regime to $q_1^2+q_2^2 \,\, \lesssim\,\, 1$ where $q_1^2+q_2^2$ is roughly characterizing the kinetic energy contribution of the initial halo system.

In measuring the spin of our halo system, we shall adopt the dimensionless spin parameter \cite{Peebles:1969jm}
\begin{align}
\lambda = \frac{J_{tot} |E_{tot}| }{  G  M^{\text{\tiny ${5}/{2}$}}_{tot}} 
\end{align}
where $J_{tot}$ denotes the magnitude of total angular momentum and $E_{tot}$ the total energy
of the system. %
%\cite{Bett:2006zy}
From the CDM-based Millennium simulation \cite{Springel:2005nw}, the average value of the spin parameter for its field halos is estimated as $\lambda_{avg}=0.0422$ \cite{Bett:2006zy}. 

In this note, we shall consider five cases %initial choices
 of $q_1=0.25,0.3,0.35,0.4,0.45$. The parameter $q_2$ will be fixed to be $q_2=0.75$. Then, for each choice of 
$q_1$ and $q_2$, we shall  adjust $Z$ to match the initial spin parameter of our initial halo configuration with the above $\lambda_{avg}$ value approximately within $5$\% errors. In this manner, one finds
$Z=90,90,90,91,91$ for $q_1=0.25,0.3,0.35,0.4,0.45$ respectively. Finally we repeat our simulation $30$ times for  each %the above five 
choice of $(q_1,q_2, Z)$ where %each set of  
initial subhalo positions and part of velocities ($\vec{v}_P$) are fully randomly generated for each simulation. Below we shall refer the resulting initial condition specified by $(q_1,q_2, Z)$ 
to ``initial condition $q_1\negthinspace$" since, with $q_1$ specified, $q_2$ and $Z$ are fully fixed in the above choices of %five choices of 
initial conditions. 
 In our CDM simulations below, these choices of initial parameters will lead to a typical CDM flattening of  satellite distributions  derived from full cosmological simulations under $\Lambda$CDM (see Section \ref{sec5} for the details).  One may try larger values of $q_1$ such as $q_1=0.8$ or even larger.  The resulting initial configurations are no longer gravitationally bound; With $q_1=0.8$, one indeed finds that part of subhalos go out of the simulation box and also that the late-time relaxation of the system is hardly achieved due to the relatively large kinetic energy. Therefore such choices will be inappropriate as our proper initial setup.

We set our total simulation time to be $%0.8\, \tau_c \simeq 
1887.0$ Myr, which corresponds to a total of 2400 time steps (in $\Delta t$).
However, we find that some of the subhalos after initial collisions may go out of the box through one side and reappear
through its opposite side, which is basically due to the periodic boundary condition imposed in this note.  
To avoid such instances with enough margin, we shall take the validity limit of our 
simulation time to be $%0.52\, \tau_c \simeq 
1226.4$ Myr. 
 %of subhalos in 
%is prepared in the following manner.
}

\def\xxx1{
In order to prepare our initial halo configuration, we divide a total of 200 initial solitons 
%(as prescribed
%in the above) \Inkyu{(어디에 설명돼 있나요?)}
into 14 groups with $n_m \, (m=1,2, \cdots, 14)$ denoting the number of initial solitons  in the $m$-th group where $(n_1, n_2, \cdots, n_{14})=(1,5,6,8,18,18, \cdots, 18)$. 
The solitons in the $m$-th group are scattered randomly over the $m$-th sphere of  radius 
$r_m =   \frac{3.78}{13} (m-1) \, {\ell_c}$, keeping at least some minimum distance between them to prevent any  significant overlap of their wave function. 
Hence the initial 200 solitons are placed within a ball of radius $R_{max}=3.78\,\ell_c\simeq 25.70\, {\rm kpc}$.
An example of such configuration is depicted on the left side of  Figure \ref{InitSats}. 

%As depicted at left side of figure \ref{InitSats}, we scattered sequentially 1, 5, 6, 8, 18, 18, ... , 18, in total 200 identical initial solitonic subhalos on spherical surfaces of uniformly increasing radial distances from 0 to $R_{max} \sim 25$, to be spherically arranged from the center of mass of them in periodic cubic box of a side length $L_{box}$ = 204 kpc ($\sim 30\, \l_c$), keeping minimum distance between subhalos for the significant overlap of them not to occur.  
To investigate the influence of the initial conditions we vary initial velocities of subhalos.  
For each subhalo (roughly referring to an initial soliton and its time-evolved configuration in the FDM case) 
located at $(r,\theta,\phi)$ in spherical polar coordinates, we assign an initial velocity %(IV)
\be
\vec{v}_0 =V(r) %\sqrt{\frac{GM_{tot}(r)}{r}}
\left(\sin\theta\,\hat\phi-|\cos\theta|\,\frac{\alpha\,+\, \beta\sin^2\theta}{1\,+\,\beta}\,\hat r\right),%\quad \alpha,\beta \in [0, 0.5, 1] 
\label{initvelocity condition}
\ee
where $V(r)$ is the rotation velocity $\sqrt{\frac{G M_{tot}(r)}{r}}$ with $M_{tot}(r)$ denoting the sum of masses within the radius $r$ from the center of the box and $\alpha,\beta=0,0.5,1$.
In the case of $\alpha=\beta=0$, all initial subhalos possess only an azimuthal component of their initial velocities, and in the remaining  eight pairs of $(\alpha, \beta)$, the initial velocity of the subhalos includes  an additional radial component $v_r = \hat{r}\cdot \vec{v}_0$.
For the total nine cases of $(\alpha,\beta)$, the corresponding angular functions 
% depending on their angular positions as
 are illustrated on the  right side of Figure \ref{InitSats}. 
Additionally, we  adjust uniformly the positions  and the velocities of all the subhalos such that the resulting halo system  has a vanishing center of mass position and velocity involving only a $z$ component of total angular momentum. % in the box coordinate system. 

\begin{figure}[htbp]   
%\vskip-3cm
\begin{center}
\includegraphics[width=0.4\textwidth]{InitSats.png}
\includegraphics[width=0.4\textwidth]{vel_a_b.png}
%\vskip-1cm
\caption{\footnotesize 
%Initial positions of subhaloes (left) and angular dependence of their radial velocities in velocity condition $\ref{initvelocity condition}$ (right). In right, the angular dependencies of radial velocities of subhalos are represented with the ratio of magnitude of radial velocity to rotational velocity,  $v_r/V(r)=\cos\theta\,\frac{\alpha\,+\, \beta\sin^2\theta}{1\,+\,\beta}$ in which $V(r)=\sqrt{\frac{GM_{tot}(r)}{r}}$ and $\theta$ is the angle between z-axis and position vector of subhalo.
An example of initial subhalo configuration in the position space is depicted on the left side. On the right hand side, we draw the angular dependence of $-v_r/V(r)$ for the total nine %$9$ 
cases of $(\alpha,\beta)$. 
}
\label{InitSats}
\end{center}
\end{figure} 

%While the total time duration we set for the system evolution is $0.8\, \tau_c \sim 1886.7$ Myr, the valid time for the analysis of results is about 50\% of it, because the subhalos go out through one side of box and come in through the opposite side roughly at that time. We repeated the simulation 30 times for each velocity condition with randomly perturbed initial positions of subhalos from the arrangement of the subhalos which led to the most proper result for our study  among many configurations we tried. When we perturb the positions of subhalos, we moved only the angular positions of them within $\pm 0.1246$ keeping the radial distance.  

We set our total simulation time to be $0.8\, \tau_c \simeq 1887$ Myr, which corresponds to a total of 2400 time steps (in $\Delta t$).
However, we find that some of the subhalos after initial collisions may go out of the box through one side and reappear
through its opposite side, which is basically due to the periodic boundary condition imposed in this note.  
To avoid such instances with enough margin, we shall take the validity limit of our 
simulation time to be $0.52\, \tau_c \simeq 1226$ Myr. 
We repeat our simulation $30$ times for each choice of $(\alpha,\beta)$ where each set of initial subhalo positions %of subhalos in 
is prepared in the following manner; Among many trials of the above prescription, we first choose a reference set of initial subhalo positions by singling out a set which shows   relatively  typical performances. With this reference set, we generate $30$ sets of initial subhalo positions where, in each set, we perturb angular positions of all subhalos randomly 
within an angular size $0.1246$ while keeping their radial positions intact.

%
% IC Co-rotation
%

Finally note that, with ($\ref{initvelocity condition}$), all subhalos in each initial velocity condition rotate initially 
in the same angular direction $\hat\phi$, which is designed %mainly in order 
to focus 
on the comparison of resulting
flattened structures. % in  resulting satellite systems. 
Our last initial velocity condition referred to ``corotation'' is designed to see the difference in the corotation ratio $\eta=\frac{N^+_s}{N_s}$ where $N^\pm_s$ denotes the number of satellites involving a positive/negative $\hat\phi$-component respectively.
The corresponding initial velocity condition is set in the following manner; We begin with the $\alpha=\beta=1$ velocity condition, randomly select $50$ initial subhalos  out of $200$ and change their initial angular direction from $\hat\phi$ to $-\hat\phi$ while reducing its angular speed by one half factor. 
Thus one has
\begin{align}
\vec{v}^{\,\pm}_0 &=V(r)%\sqrt{\frac{GM_{tot}(r)}{r}}
\left( \frac{1}{4}(1\pm 3)\,\sin\theta \, %-\frac{\sin\theta}{2}\,
\hat\phi-|\cos\theta|\,\frac{1\,+\, \sin^2\theta}{2}\,\hat r\right)\ , %\nonumber\\ 
%\vec{v}^{\,+}_0 &=V(r)% \sqrt{\frac{GM_{tot}(r)}{r}}
%\left(\sin\theta\,\hat\phi-\cos\theta\,\frac{1\,+\, \sin^2\theta}{2}\,\hat r\right)\ ,
\label{corotationvelocity condition}
\end{align}
as their initial velocity condition. We shall also use this corotation velocity condition to see differences in 
the above measures %in the above 
as it works as a more general form of initial velocity condition.
%\subsection{Co-rotation}\label{sec2.3}
%All satellites of simulations of our halo models had been initialized to rotate in same direction ($\ref{initvelocity condition}$) for the comparison of the flattening of those systems.
%We compared corotation of satellites in our halo models with ratio ($\eta = \frac{N^+_s}{N_s}$) of number of corotating satellites($N^+_s$) to that of all satellites ($N_s = N^+_s + N^-_s $). For this corotation test, we set randomly selected 25 percent of initial satellites or subhalos ($N^-_s=50$) to rotate in opposite direction with a half speed of that in former simulations with $\alpha=\beta=1$, so that 50 subhalos rotating in $- \phi$ direction and 150 subhalos in $+ \phi$ direction have their corresponding initial velocities 
%\begin{align}
%\vec{v}^{\,-}_0 &= \sqrt{\frac{GM_{tot}(r)}{r}}\left(-\frac{\sin\theta}{2}\,\hat\phi-\cos\theta\,\frac{1\,+\, \sin^2\theta}{2}\,\hat r\right),\\
%\vec{v}^{\,+}_0 &= \sqrt{\frac{GM_{tot}(r)}{r}}\left(\sin\theta\,\hat\phi-\cos\theta\,\frac{1\,+\, \sin^2\theta}{2}\,\hat r\right).
%\label{corotationvelocity condition}
%\end{align}
}

\def\xxx2{
\subsection{Measures of satellite distributions }\label{subsec:measures}
The typical spatial distributions of satellites in our FDM halo systems show rather clear differences
from those of the CDM side (See Figures \ref{5x4DFmap} and \ref{mgsDFmap}). In addition the typical numbers of satellites in the FDM side are only about 25 percent of those in the CDM side (See again Figures \ref{5x4DFmap} and \ref{mgsDFmap}). % Below %To see these differences clearly, 
We shall use the following
measures in order to demonstrate that the FDM halo systems have %a tendency to form 
more flattened 
 satellite structures %of satellites
than their CDM counterparts. 

First let us introduce the ratio of the semi-minor to the semi-major axis (denoted as $c/a$) of an ellipsoid derived
from a satellite distribution  \cite{Shao:2019nuc, Shao:2016nsx, Zentner:2005wh}  in the following manner. Its principal axes are defined using the mass tensor
of satellites
\begin{equation}\label{Iij}
I_{ij} \equiv \frac{1}{N_s}\sum^{N_s}_{k=1} x^{(k)}_{i} \, x^{(k)}_{j}
\end{equation}
where ${\vec{x}}_{(k)}$ refers to the position %vector 
of the $k$-th satellite and $N_s$
denotes the total number of satellites in each halo system.  Note here that the origin of the position is defined by the center of mass in each halo system where its density peak is also located approximately. In fact, with our choice of initial halo configuration, this center of mass was set to agree with the center of the simulation box, and shall be simply called the center below. 
%for simplicity. 
The eigenvalues of the tensor (ordered by $\lambda_1 \ge \lambda_2 \ge \lambda_3$) define
the corresponding principal semi-axes respectively by $a=\sqrt{\lambda_1}$, $b=\sqrt{\lambda_2}$, and $c=\sqrt{\lambda_3}$.

%Adopting a method in other related works, we calculated the ratios of minor to major axis ($c/a$) of ellipsoids of the satellites in each CDM and FDM halo model.
%The principal axes of the corresponding ellipsoids are calculated with following mass tensor of satellites,
%
%where $x_{k,i}$ represents the i-th component of the position vector for satellite k with respect to the center of host halo which is defined with the peak of the largest among the clustered by halo finder. The eigen values $\lambda_i$ of the mass tensor give the three principal axes $a=\sqrt{\lambda_1}$, $b=\sqrt{\lambda_2}$, and $c=\sqrt{\lambda_3}$ in descending order. 

%As seen in figure \ref{2x2DFmap} or figure \ref{mgsDFmap}, there are much less number of satellites in the FDM halo than in the CDM. If there are not enough satellites within small radius from the center of FDM host halo, $c/a$ for the satellites within that radius cannot be calculated or biased to be smaller than for that in CDM halo. So $c/a$ is not adequate for comparison of the profile along radius $r$.  

An alternative to the ratio $c/a$ is the $r$-dependent cumulative distribution function (CDF) $f_1(|\cos \theta|;r)$\footnote{Slightly abusing notation, 
we shall denote the full CDF including the whole satellites simply by $f_1(|\cos \theta|)$.} of
a satellite system where one includes only satellites within 
the radius $r$ from the center and $\cos\theta$ is given by $\hat{e}_3 \cdot \hat{r}$ with $\vec{r}$ being the position of a satellite and $\hat{e}_3$ denoting the unit eigenvector corresponding to the eigenvalue $\lambda_3$. Specifically, we introduce a quantity $\A(<r)$ defined by
\begin{equation}\label{avr_cos_theta}
\A(<r) \equiv 
\frac{1}{N_s(r)}\sum^{N_s(r)}_{k=1} |\cos\theta_{(k)}|  \,,
%\frac{1}{N_s}\sum^{N_s}_{k=1}\cos\theta_k
\end{equation}
where we again include only 
%$N_s(r)$ denotes the number of
satellites within the radius $r$ from the center and $N_s(r)$ denotes their number.
This equals to the remaining area above the CDF curve in Figure \ref{CDFcos_plot}, which 
may be shown as
\begin{equation}\label{avr1_cos_theta}
\A(<r)  =\int_0^1 x\,\frac{df_1(x;r)}{dx}\, dx = 1 -  \int_0^1 f_1(x;r)dx \ .
%\frac{1}{N_s}\sum^{N_s}_{k=1}\cos\theta_k
\end{equation}
We shall also use the Kolmogorov-Smirnov probability $P_{ks}$ by comparing CDF samples of the FDM/CDM
satellite systems where each CDF denoted by $f_{30}(|\cos \theta|)$ is drawn from  the angular positions of satellites in all 30 samples for each choice of initial velocity condition.
(In general this quantity represents the probability that any two sets of samples were drawn from the same cumulative probability distribution.)
With the CDF's as well as
$P_{ks}$, we may deduce the main underlying factors that make the FDM halo system have a more flattened
satellite system than its CDM counterpart.

The last measure $\delta$ is the  rms of vertical distances of satellites from their best-fit-plane \cite{Zentner:2005wh}
that is determined by minimizing 
\begin{equation}\label{delta-1}
\delta \equiv \sqrt{ \frac{\sum^{N_s}_{k=1} (\hat{n}\cdot\vec x_{(k)})^2 }{N_s} }
\end{equation}
with respect to $\hat{n}$ where $\hat{n}$ is a unit normal vector to the would-be best-fit-plane. In other words the best-fit-plane is obtained with all satellites in the whole simulation box.
In fact the corresponding unit normal vector $\hat{n}$ determined in this manner almost agrees with the $\hat{e}_3$
eigenvector given %in the 
above. One further introduces the rms height $\delta(<\rho)$ calculated only with the satellites within the {\it horizontal} distance $\rho$ from the center, 
%of the halo,
%is less than $\rho$ 
in which the best-fit-plane and unit normal vector  $\hat{n}$ %have already been 
are in advance fixed by the minimization procedure of $\delta$ in (\ref{delta-1}).

%\begin{equation}\label{delta}
%\delta \equiv \sqrt{\frac{\sum^{N_s}_{k=1} (\hat{n}\cdot\vec x_k)^2  }{N_s}}
%\end{equation}
%where $\vec x_k$ represents the position vector for satellite k with respect to the center of host halo and $\hat{n}$ is unit normal vector of best-fit plane. In fact, the unit normal vector to the plane determined by above method closely correspond to the minor axis ($\vec e_3$) of ellipsoid of satellites in the first measure. 

}

%-------------------------------------------------------------------------------
\subsection{Halo reconstruction}\label{sec:mgs}
%-------------------------------------------------------------------------------
Given the initial halo configuration, the simulations using PyUltraLight for FDM and Gadget2 for CDM are performed.
For each time step, we save both FDM and CDM density maps for further halo reconstruction and data analysis.
%The first step of the analysis is to find cluster structures in the simulated data.
Galaxies and halos are the gravitationally bound objects and they appear as clusters in the density map data. 
A cluster can be defined as a large collection of points or cells in a small area or a localized volume.
Cluster finding has been a popular topic in computational science. 
Many different clustering algorithms are available which have been developed and maintained by many different groups.
As each clustering algorithm has its own strength and weakness, one should carefully choose the right algorithm for their own research purpose. 

In this study, we have adopted the Mulguisin clustering algorithm, which has been used in particle physics for finding jet structures from the collection of particle tracks \cite{mulguisin}.
The algorithm was modified to find clusters out of input cells in 3D space.
It first finds the most massive cell in the input data and name it a seed.
Then it looks for the next massive cell which will be used as a test cell.
%A test cell is to be attached to the seed according to a certain condition.
%We use the distance between two cells as the condition.
If the test cell is close enough to the seed, we then attach the cell to the seed and these two cells become a group.
A group can therefore be regarded as a list of cells. 
If the distance between the test cell and the seed is longer than a certain cut length, then it becomes a new seed.
In this study we use 1.4 kpc as out distance cut length.

Now the same procedure is repeated for the next massive cell.
The cell will be attached to a group when the minimum distance between the test cell and the group is shorter than the cut length.
For this we calculate all distances between the test cell and the cells in the group and choose the minimum value. 
Once again, if the test cell does not join to any existing groups then it becomes a new seed.
The algorithm keeps doing this simple process until there are no test cells left.
At this stage, cells are all converted to groups.
Some groups are big, {\it i.e.} have many cells, and some are not.
There are also single-cell groups that are isolated from other groups.  

%\begin{figure}[tbp]   
%\begin{center}
%\includegraphics[width=0.9\textwidth]{mgs-centroid.png}
%\caption{\footnotesize (a) Reconstructed centroid as a super-cluster (central black dots) and  clusters %(subhalos) 
%around the centroid. (b) Cluster %Subhalo 
%distribution after removing the centroid.}
%\label{mgs-centroid}
%\end{center}
%\end{figure} 

At this stage a group is called a $cluster$ when its mass is larger than a certain mass cut.
Clusters are then further tested to see whether they form a super-cluster.
A super-cluster can therefore be regarded as a pack of clusters that are attached with each other.
The minimum distance between two clusters is calculated from the whole combinations of cells from each cluster and it is used to decide whether the two clusters are to be merged or not.
We use the same distance cut of $1.4$ kpc for the merging process.

All these procedures are implemented in our Mulguisin halo finder software.
%Figure \ref{mgs-centroid}{\color{blue}(a)} shows the cluster distributions that are found in a typical CDM data using the Mulguisin halo finder.
%One can see the algorithm successfully finds the centroid (central galaxy in our context) %structure as a super-cluster which is a pack of clusters stuck together.
%Figure \ref{mgs-centroid}{\color{blue}(b)} simply shows the cluster distribution after removing the centroid.
Figure \ref{mgsDFmap} shows the results from cluster finding by the Mulguisin algorithm.
The clusters that were found by the halo finder are marked with black circles.

%-------------------------------------------------------------------------------
\section{Numerical Results}\label{sec4}

\begin{figure}[htbp]   
%\vskip-3cm
\begin{center}
\includegraphics[width=1.0\textwidth]{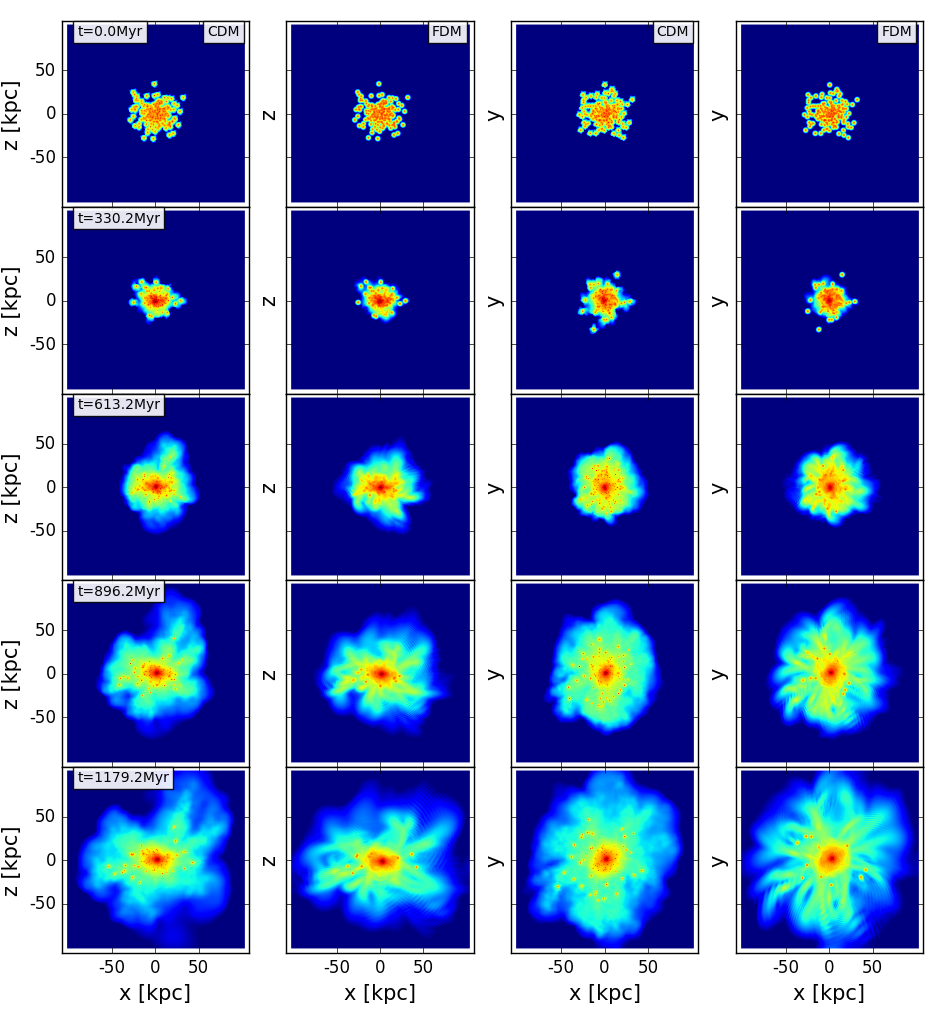}
%\vskip-1cm
\caption{\footnotesize The time evolution of halos in the CDM and the FDM models. The  left two columns are for the projected maps in the $x$-$z$ plane and the right two %columns in 
in the $x$-$y$ plane with  { the $q_1=0.35$ initial condition.} 
%\sout{the initial velocity condition %(\ref{corotationvelocity condition}).} 
%The first column shows the evolution of CDM halo, the second of FDM halo, the third of CDM halo, and the fourth of FDM halo.
The first and  third column show  the time evolution of the CDM halos while the remaining
are for the evolution of the FDM halos. For these projected maps,
the mass density fields of each halo system are { integrated} along $y$-direction (the left two columns) and $z$-direction (the right two columns) of the full simulation box. They %The projected maps 
are (decimal-)log-scaled and truncated with {max $= -0.778$ and min $= -6.86$.}
% for the first three  rows and max $= -0.434$ and min $= -6.51$ for the remaining.
}
\label{5x4DFmap}
\end{center}
\end{figure}

\begin{figure}[htbp]   
%\vskip-0.2cm
\begin{center}
\includegraphics[width=0.95\textwidth]%{mgsSats.png}
{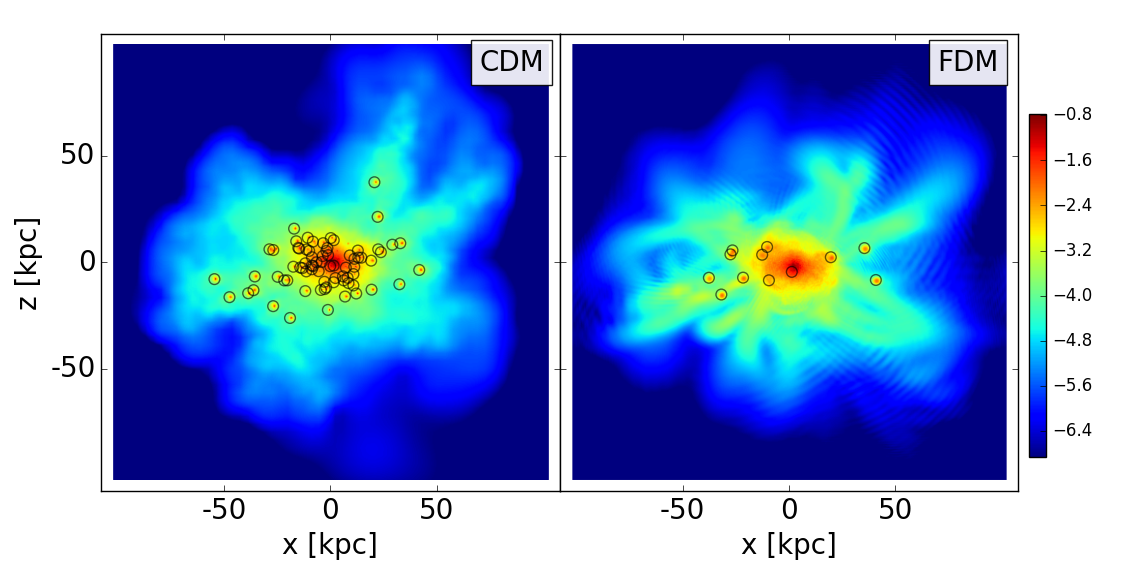}
%\vskip+0.2cm
\caption{\footnotesize %Satellites in the 
Projected maps in the $x$-$z$ plane and
their satellites are illustrated in this figure. %along $y$ direction % $x$-$z$ plane of the simulation box. 
To make the maps, we used the snapshots at $t= 1179.2$ Myr
%of samples 
from the simulations 
for the corotation test. 
%with the initial velocity condition (\ref{corotationvelocity condition}).
Note that the two maps basically coincide  with the ones in the last row of the left two columns
in Figure \ref{5x4DFmap}.  
They %The projected maps 
are  (decimal-)log-scaled and truncated with { max $= -0.778$ and min $= -6.86$}.
%on the same projected maps shown in Figure %\ref{2x2DFmap}.
In these maps, the satellites 
%(right column) 
 are marked with black circles, which are identified by the Mulguisin halo finder.
}
\label{mgsDFmap}
\end{center}
\end{figure}

\begin{figure}[htbp]   
%\vskip-1cm
\begin{center}
\includegraphics[width=1.0\textwidth]{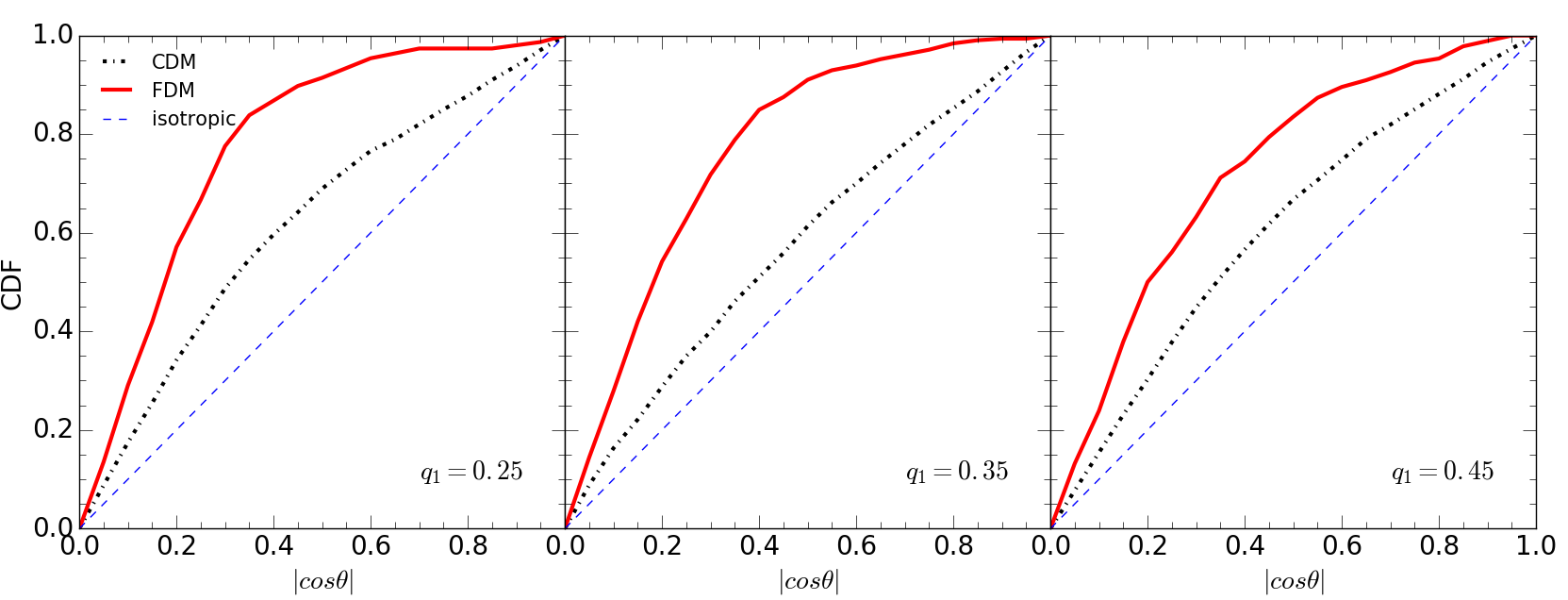}
\vskip-0.2cm
\caption{\footnotesize The 
%cumulative distribution function (CDF),
CDF $f_{30}(|\cos\theta|)$ is depicted in this figure. 
%of angular positions of satellites. 
In each panel, the  black dash-dotted line is for the CDM, the red solid line for the FDM, and the blue dashed  corresponds to %homogeneous 
the reference spherically symmetric
distribution. The distribution of $|\cos\theta|$ 
%of satellites 
%from the simulations 
with each initial %velocity 
condition is built
by  collecting 
all %$|\cos\theta|$'s of 
satellites from each set of %corresponding 
30 samples (see Section \ref{sec2.1.1}) at $t= 1179.2$ Myr.}
\label{CDFcos_plot}
\end{center}
\end{figure}

\def\qqq8{
\subsection{Measures of satellite distributions }\label{subsec:measures}
The typical spatial distributions of satellites in our FDM halo systems show rather clear differences
from those of the CDM side (See Figures \ref{5x4DFmap} and \ref{mgsDFmap}). In addition the typical numbers of satellites in the FDM side are only about 25 percent of those in the CDM side (See again Figures \ref{5x4DFmap} and \ref{mgsDFmap}). % Below %To see these differences clearly, 
We shall use the following
measures in order to demonstrate that the FDM halo systems have %a tendency to form 
more flattened 
 satellite structures %of satellites
than their CDM counterparts. 

First let us introduce the ratio of the semi-minor to the semi-major axis (denoted as $c/a$) of an ellipsoid derived
from a satellite distribution  \cite{Shao:2019nuc, Shao:2016nsx, Zentner:2005wh}  in the following manner. Its principal axes are defined using the mass tensor
of satellites
\begin{equation}\label{Iij}
I_{ij} \equiv \frac{1}{N_s}\sum^{N_s}_{k=1} x^{(k)}_{i} \, x^{(k)}_{j}
\end{equation}
where ${\vec{x}}_{(k)}$ refers to the position %vector 
of the $k$-th satellite and $N_s$
denotes the total number of satellites in each halo system.  Note here that the origin of the position is defined by the center of mass in each halo system where its density peak is also located approximately. In fact, with our choice of initial halo configuration, this center of mass was set to agree with the center of the simulation box, and shall be simply called the center below. 
%for simplicity. 
The eigenvalues of the tensor (ordered by $\lambda_1 \ge \lambda_2 \ge \lambda_3$) define
the corresponding principal semi-axes respectively by $a=\sqrt{\lambda_1}$, $b=\sqrt{\lambda_2}$, and $c=\sqrt{\lambda_3}$.

%Adopting a method in other related works, we calculated the ratios of minor to major axis ($c/a$) of ellipsoids of the satellites in each CDM and FDM halo model.
%The principal axes of the corresponding ellipsoids are calculated with following mass tensor of satellites,
%
%where $x_{k,i}$ represents the i-th component of the position vector for satellite k with respect to the center of host halo which is defined with the peak of the largest among the clustered by halo finder. The eigen values $\lambda_i$ of the mass tensor give the three principal axes $a=\sqrt{\lambda_1}$, $b=\sqrt{\lambda_2}$, and $c=\sqrt{\lambda_3}$ in descending order. 

%As seen in figure \ref{2x2DFmap} or figure \ref{mgsDFmap}, there are much less number of satellites in the FDM halo than in the CDM. If there are not enough satellites within small radius from the center of FDM host halo, $c/a$ for the satellites within that radius cannot be calculated or biased to be smaller than for that in CDM halo. So $c/a$ is not adequate for comparison of the profile along radius $r$.  

An alternative to the ratio $c/a$ is the $r$-dependent cumulative distribution function (CDF) $f_1(|\cos \theta|;r)$\footnote{Slightly abusing notation, 
we shall denote the full CDF including the whole satellites simply by $f_1(|\cos \theta|)$.} of
a satellite system where one includes only satellites within 
the radius $r$ from the center and $\cos\theta$ is given by $\hat{e}_3 \cdot \hat{r}$ with $\vec{r}$ being the position of a satellite and $\hat{e}_3$ denoting the unit eigenvector corresponding to the eigenvalue $\lambda_3$. 
\def\qqq3{Specifically, we introduce a quantity $\A(<r)$ defined by
\begin{equation}\label{avr_cos_theta}
\A(<r) \equiv 
\frac{1}{N_s(r)}\sum^{N_s(r)}_{k=1} |\cos\theta_{(k)}|  \,,
%\frac{1}{N_s}\sum^{N_s}_{k=1}\cos\theta_k
\end{equation}
where we again include only 
%$N_s(r)$ denotes the number of
satellites within the radius $r$ from the center and $N_s(r)$ denotes their number.
This equals to the remaining area above the CDF curve in Figure \ref{CDFcos_plot}, which 
may be shown as
\begin{equation}\label{avr1_cos_theta}
\A(<r)  =\int_0^1 x\,\frac{df_1(x;r)}{dx}\, dx = 1 -  \int_0^1 f_1(x;r)dx \ .
%\frac{1}{N_s}\sum^{N_s}_{k=1}\cos\theta_k
\end{equation}
}
We shall  use the Kolmogorov-Smirnov probability $P_{ks}$ by comparing CDF samples of the FDM/CDM
satellite systems where each CDF denoted by $f_{30}(|\cos \theta|)$ is drawn from  the angular positions of satellites in all 30 samples for each choice of initial velocity condition.
(In general this quantity represents the probability that any two sets of samples were drawn from the same cumulative probability distribution.)
With the CDF's as well as
$P_{ks}$, we may deduce the main underlying factors that make the FDM halo system have a more flattened
satellite system than its CDM counterpart.

The last measure $\delta$ is the  rms of vertical distances of satellites from their best-fit-plane \cite{Zentner:2005wh}
that is determined by minimizing 
\begin{equation}\label{delta-1}
\delta \equiv \sqrt{ \frac{\sum^{N_s}_{k=1} (\hat{n}\cdot\vec x_{(k)})^2 }{N_s} }
\end{equation}
with respect to $\hat{n}$ where $\hat{n}$ is a unit normal vector to the would-be best-fit-plane. In other words the best-fit-plane is obtained with all satellites in the whole simulation box.
In fact the corresponding unit normal vector $\hat{n}$ determined in this manner almost agrees with the $\hat{e}_3$
eigenvector given %in the 
above. One further introduces the rms height $\delta(<\rho)$ calculated only with the satellites within the {\it horizontal} distance $\rho$ from the center, 
%of the halo,
%is less than $\rho$ 
in which the best-fit-plane and unit normal vector  $\hat{n}$ %have already been 
are in advance fixed by the minimization procedure of $\delta$ in (\ref{delta-1}).

%\begin{equation}\label{delta}
%\delta \equiv \sqrt{\frac{\sum^{N_s}_{k=1} (\hat{n}\cdot\vec x_k)^2  }{N_s}}
%\end{equation}
%where $\vec x_k$ represents the position vector for satellite k with respect to the center of host halo and $\hat{n}$ is unit normal vector of best-fit plane. In fact, the unit normal vector to the plane determined by above method closely correspond to the minor axis ($\vec e_3$) of ellipsoid of satellites in the first measure. 

}

In this section we shall present the outcome of our numerical study. % are presented.
In Figure \ref{5x4DFmap} we depict the time evolution of DM %dark matter 
density during the 
 formation of sample galactic halos % systems
 (either in the  CDM or the FDM model)
 as a time series of projected DM %dark matter 
 density  maps in the unit of { $10^{10} M_\odot/\text{kpc}^2$.}
 To make these projected maps, the mass density fields of each halo system
are %\sout{fully added}
{ integrated over the range $[-L_{box}/2, L_{box}/2]$}
along $y$-direction (the left two columns) or $z$-direction (the right two columns) of the
simulation box.
 %or $z$-direction  of the simulation box (in the figure, the first two columns are for $x$-$z$ planes while the last two for $x$-$y$ planes). 
%\sout{These maps %The projected maps 
 %are  (decimal-)log-scaled and truncated with the maximum $= -0.778$ and the minimum $= -6.86$.}
 % for the first three  rows and max $= -0.434$ and min $= -6.51$ for the remaining. 
 %in the above units.
 Also for these maps, we used the samples from 
%\sout{the simulations for the corotation %test with the initial velocity condition %(\ref{corotationvelocity condition}).
%} 
{ the simulations for the $q_1=0.35$ initial condition.}
Note that both the CDM and the FDM halo density maps shown in the figure are from %those 
 simulations with the same initial positions and velocities of subhalos.

% or -15 (-22 for visual identification of gravitational cooling in second and third rows, -15 for sharp contrast of halos to its background in fourth and fifth rows). 

  The CDM and the FDM halo system %in both the CDM and the FDM 
  look quite similar  at the early stages of the evolution %formation 
  process. But after the collapse  of initial subhalo systems
  %\sout{which happens mainly along $z$-direction} 
  as illustrated in the second row of Figure \ref{5x4DFmap}, their difference %between the CDM and the FDM system 
  becomes clear as will be further described below: 
 In the CDM side, there appear far more surviving subhalos, especially around the central region, than the FDM counterpart, and so is more substructure on the smallest scales.
 %Above
 %all the FDM profiles look 
% bulky compared to their CDM counterparts.   
 On the length scales of the whole parent halo, however, the tentacle-like structure found at late times 
is more prominent in the FDM  than the CDM side.

\def\qqqqq{{
\begin{figure}[htbp]   
%\vskip-3cm
\begin{center}
\includegraphics[width=0.9\textwidth]{corotation3to1Vphi05_T05_2D_seed9.png}
%\vskip-1cm
\caption{\footnotesize Projected maps of mass density fields of the CDM and the FDM halos. 
For these projected maps,
the mass density fields of each halo system are fully added along $z$-direction (upper panels) and $y$-direction (lower panels) of the simulation box. They %The projected maps 
are log-scaled and truncated with max $= -1.0$ and min $= -15$. To make them %these maps, 
we used the snapshots at $t=0.5\, \tau_c \,(\sim 1179.2$ Myr) of samples from the simulations for corotation test with the initial velocity condition (\ref{corotationvelocity condition}). Note that both the CDM and the FDM %halo 
density fields are from %results of
the simulations with the same initial positions and velocities of subhalos.}
\label{2x2DFmap}
\end{center}
\end{figure} 
}}

% The snapshots in Figures \ref{2x2DFmap} and \ref{mgsDFmap} more clearly show the difference in the late time distributions of satellites in our CDM and FDM models in the bottom row of the Figure \ref{5x4DFmap}. 
% In Figure \ref{mgsDFmap}, one finds that the FDM system has a more planar distribution and a smaller number %less
%  of satellite galaxies  %systems
% than the CDM counterparts.
 %The satellites marked with small circles on the projection density map in the right column of Figure \ref{mgsDFmap} are objects with mass greater than $2\times 10^7M_{\odot}$ among the clustered by Mulguisin algorithm described in Section \ref{sec3}. 
Among clusters found by the Mulguisin algorithm, 
%\sout{described in Section \ref{sec3},} 
we shall identify objects with mass greater than $2\times 10^7M_{\odot}\,$\footnote{This mass-cut %scale 
is introduced to eliminate  too small clusters such as globular clusters.}  as satellites,  which were all marked with small black circles on the projected maps in %\sout{the right column of} 
Figure \ref{mgsDFmap}. 
%\sout{In this figure, the left two maps %coincide  with the ones in the last row of %the left two columns
%in Figure \ref{5x4DFmap}.} % They 
%Note that these maps 
%are  (decimal-)log-scaled and truncated with max {\color{blue}$= -0.778$ and min $= -6.86$.}
We find that %In this case
the CDM system has roughly {eight} times more satellites than the FDM side. While there are many satellites in the central region of the CDM halos, the FDM halos involve 
almost none in that region. One may also see that the FDM satellite distributions %in the FDM
appear more flattened than  the CDM counterparts, %as is clear 
even before analyzing any details. %\sout{the measures for the degree of %flattening %described 
%in Section %\ref{subsec:measures}.} 
One can also see the outgoing waves from the 
gravitational cooling in the FDM side.

\begin{figure}[htb]   
\vskip-0.1cm
\begin{center}
\includegraphics[width=1.0\textwidth]{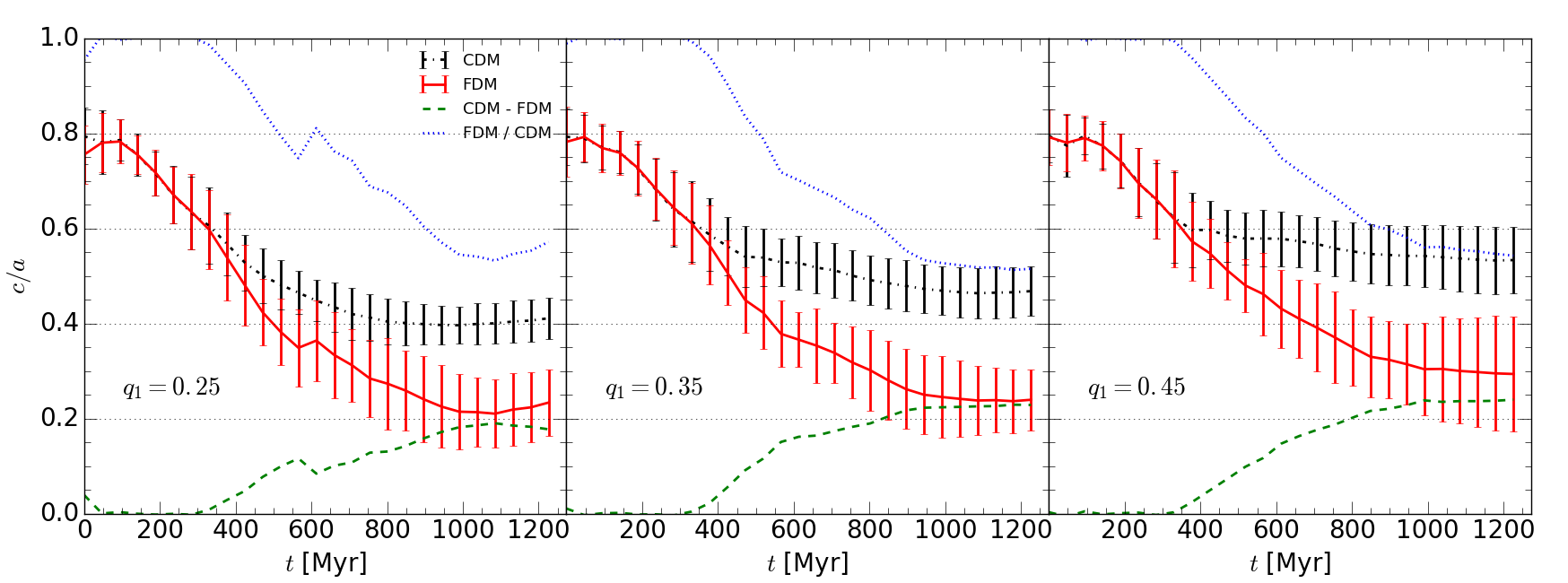}
\vskip-0.2cm
\caption{\footnotesize The evolution of $c/a$ from the initial time $t=0$ to { $1226.4$ Myr for %the satellites from 
the initial conditions %simulations 
with %$\alpha=\beta=0$ (top left), $\alpha=1,\beta=0$ (top right),  $\alpha=\beta=1$ (bottom left) 
$q_1=0.25$ (left), $0.35$ (middle) and $0.45$ (right).
%of the initial velocity condition (\ref{initvelocity condition}) 
%and %the satellites from the simulation for
%the corotation %test with initial velocity condition 
%(\ref{corotationvelocity condition}) (bottom right). 
}
In each panel, the black dash-dotted line shows the ratio $\langle c/a \rangle$ of satellites in the CDM halo model, the red solid that in the FDM, the green dashed line %shows 
their difference, $\langle (c/a)_{\mbox{\tiny CDM}}%\negthinspace
-%\negthinspace  
(c/a)_{\mbox{\tiny FDM}} \rangle$, and the blue dotted  their ratio, $\langle (c/a)_{\mbox{\tiny FDM}} /  (c/a)_{\mbox{\tiny CDM}} \rangle$.   The error bars in each plot represent the estimation of 
%standard error of the mean (
{ standard deviation.}% with 
%number of samples 
%$n=30$.
%, here $\sigma$ is the sample standard deviation on $c/a$. 
}
\label{c/a_plot}
\end{center}
\end{figure}

\begin{figure}[htbp]   
%\vskip-0.2cm
\begin{center}
\includegraphics[width=1.0\textwidth]{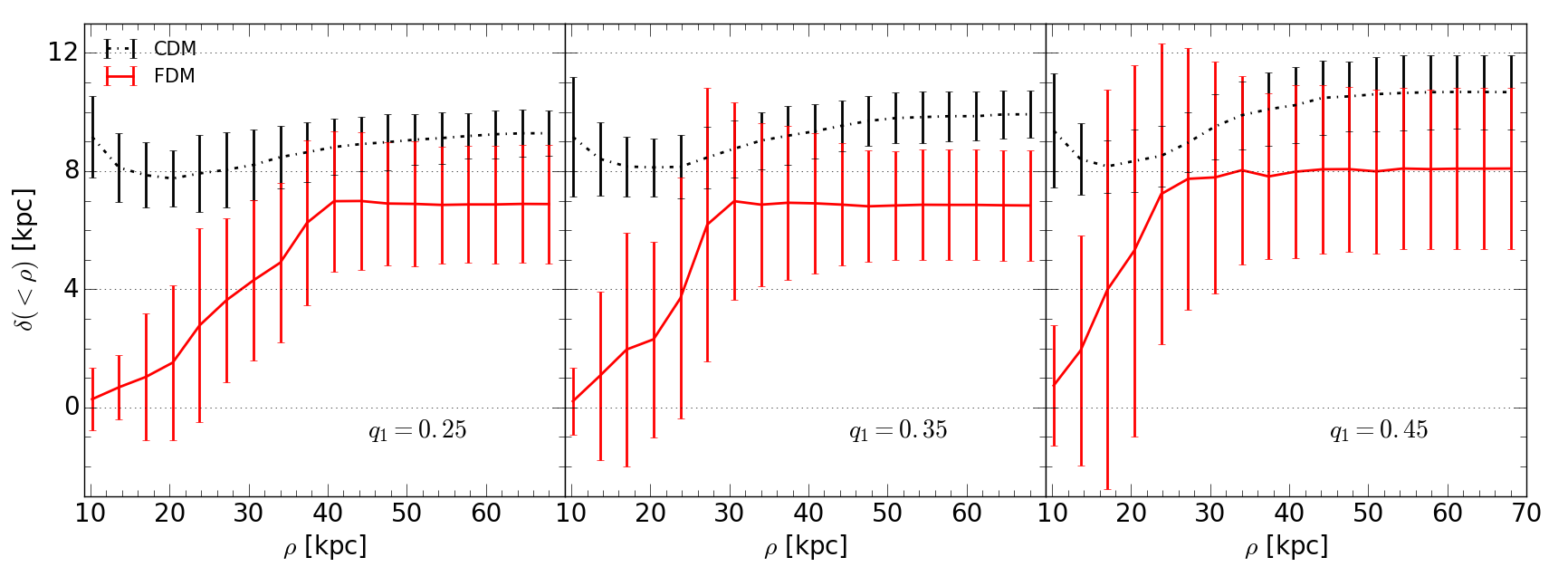}
\vskip-0.2cm
\caption{\footnotesize The rms 
height %of perpendicular distances 
$\delta (<\rho)$ of satellites from the reference best-fit-plane %(\ref{delta-1}) 
as a function of the distance $\rho$ from the center 
%of the halo 
at $t=1179.2$ Myr. The best-fit-plane is in advance determined with whole satellites within the full simulation box, and then the rms height $\delta(<\rho)$ is calculated only with the satellites within the horizontal distance $\rho$ from the center. % of halo. % as explained in Section \ref{sec2.3}. %is less than $\rho$.
}
\label{delta_plot}
\end{center}
\end{figure}

\begin{figure}[bhtp]   
\vskip-0.5cm
\begin{center}
\includegraphics[width=0.59\textwidth]{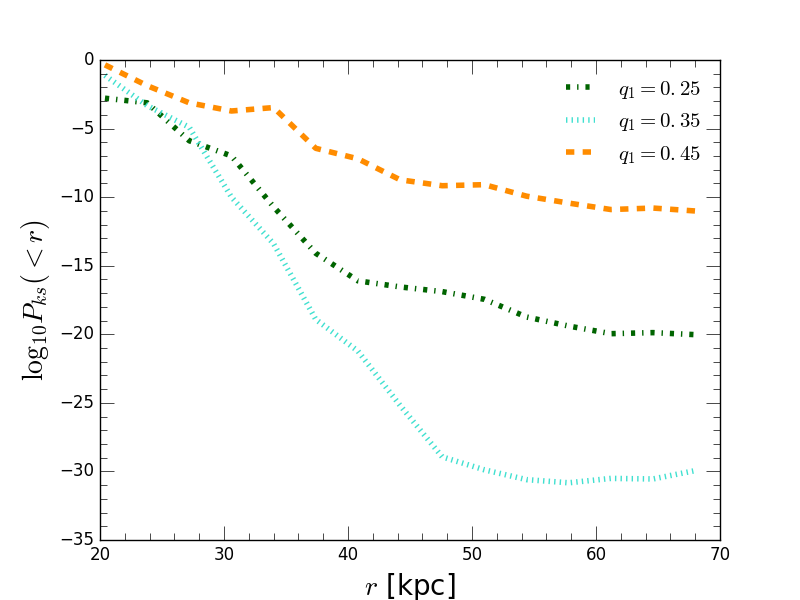}
\vskip-0.3cm
\caption{\footnotesize 
%The Kolmogorov-Smirnov probability ($P_{ks}$) for the CDFs of $|\cos\theta|$ of satellites in CDM and FDM halo models to be equivalent. This probability that two empirical distributions on satellites in CDM and FDM halo models can be drawn from the same distribution is calculated with the distributions plotted in Figure \ref{CDFcos_plot}.
The Kolmogorov-Smirnov probability $P_{ks}$ 
as a function of $r$
where only the satellites within the radius $r$ from the center 
%of halo 
are included for the evaluation
of the relevant %corresponding
CDF $f_{30}(|\cos \theta|;r)$.
%The horizontal axis in the plot represents the radius $r$ from the center of halo system, only within which the satellites are included for the evaluation of the CDFs.  
}
\label{Pks_plot}
\end{center}
\end{figure}

To be more concrete, 
%\sout{ let us now consider the measures in Section  \ref{subsec:measures}. Below} 
we shall focus on the following three choices of initial conditions, { $q_1=0.25,0.35,0.45$,}
%\sout{($\alpha,\beta$) $=(0,0),(1,0),(1,1)$
%and the corotation 
% (\ref{corotationvelocity condition}),}
particularly for 
Figures \ref{c/a_plot}--\ref{Pks_plot} and their discussions. 

{ First let us introduce the ratio of the semi-minor to the semi-major axis (denoted as $c/a$) of an ellipsoid derived
from a satellite distribution  \cite{Shao:2019nuc, Shao:2016nsx, Zentner:2005wh}  in the following manner. Its principal axes are defined using the mass tensor
of satellites
\begin{equation}\label{Iij}
I_{ij} \equiv \frac{1}{N_s}\sum^{N_s}_{k=1} x^{(k)}_{i} \, x^{(k)}_{j}
\end{equation}
where ${\vec{x}}^{(k)}$ refers to the position %vector 
of the $k$-th satellite and $N_s$
denotes the total number of satellites in each halo system.  Note here that the origin of the position is defined by the center of mass in each halo system where its density peak is also located approximately. In fact, with our choice of initial halo configuration, this center of mass was set to agree with the center of the simulation box, and shall be simply called the center below. 
%for simplicity. 
The eigenvalues of the tensor (ordered by $\lambda_1 \ge \lambda_2 \ge \lambda_3$) define
the corresponding principal semi-axes respectively by $a=\sqrt{\lambda_1}$, $b=\sqrt{\lambda_2}$, and $c=\sqrt{\lambda_3}$.}

In Figure \ref{c/a_plot}, we depict  the time evolution of $\langle c/a \rangle$,  {\it i.e.} the average of $c/a$ for the 30 %\sout{initial-position sets} 
{ simulation samples} prescribed in Section \ref{sec2.1.1}, %samples,
from the initial time $t=0$ to { $1226.4$ Myr.} %of the satellites systems 
In each plot, the black dash-dotted line is for the $\langle c/a \rangle$ of satellites system in the CDM halo model,  the red solid %line 
for that of the FDM, the green dashed %line %shows 
for their difference, 
%of $\langle c/a \rangle$ between CDM and FDM, 
and the blue dotted %line 
for the ratio of the latter over the former.
The error bars in each plot represent the estimation for the { standard deviation %error of the mean 
%(std) 
of} 
%\sout{ $\sigma/\sqrt{n}$ with %number of samples $n=30$, where $n$ denotes the number of initial-position sets 
%introduced in Section \ref{sec2.1.1} and $\sigma$ 
%is the standard deviation of $c/a$ 
%from} %with each set of the %resulting 
{ the %resulting 
30 simulation samples.} % derived.
%FDM to CDM.   
% \sout{The oscillatory behaviour of $c/a$ is due to the infall and bouncing of the subhalos during the structure formation.} 
It is clear that the FDM satellite profiles lead to a %have in general 
smaller $c/a$ than their CDM counterparts. %satellite systems.
{ In addition, for both the FDM and the CDM, one finds that  $\langle c/a\rangle$ gets larger
as $q_1$ becomes larger. This is because, as $q_1$ becomes larger, the average strength of each random velocity $\vec{v}_P$ in our initial halo configuration increases.}

%In order to examine the difference between the CDM and the FDM halo model it might be helpful to compare each  snapshot in the Figure \ref{5x4DFmap} with the value at the corresponding time in the plot for corotation test (bottom right) in the Figure \ref{c/a_plot}.
The { rms height} %last measure 
$\delta$ is the  rms of vertical distances of satellites from their best-fit-plane \cite{Zentner:2005wh}
that is determined by minimizing 
\begin{equation}\label{delta-1}
\delta \equiv \sqrt{ \frac{\sum^{N_s}_{k=1} (\hat{n}\cdot\vec x_{(k)})^2 }{N_s} }
\end{equation}
with respect to $\hat{n}$ where $\hat{n}$ is a unit normal vector to the would-be best-fit-plane. In other words the best-fit-plane is obtained with all satellites in the whole simulation box.
In fact, the corresponding unit normal vector $\hat{n}$ determined in this manner almost agrees with the $\hat{e}_3$
eigenvector given %in the 
above. One further introduces the rms height $\delta(<\rho)$ calculated only with the satellites within the {\it horizontal} distance $\rho$ from the center, 
%of the halo,
%is less than $\rho$ 
in which the best-fit-plane and unit normal vector  $\hat{n}$ %have already been 
are in advance fixed by the minimization procedure of $\delta$ in (\ref{delta-1}).
 In Figure \ref{delta_plot}, we depict the rms
height %of perpendicular distances 
$\delta(<\rho)$ of satellites 
%to the best-fit-plane (\ref{delta-1}) 
as a function of the horizontal distance $\rho$ from the center 
%of halo 
at $t= 1179.2$ Myr,  where we have taken $\rho \ge 10$ kpc because we would like to avoid the centroid residing in the region within $10$ kpc.  
%The relevant 
%best-fit-plane here is obtained with all satellites within the full %entire
%simulation box, and then 
We may confirm that the FDM systems have thinner satellite planes than the CDM counterparts.  In particular, one finds that, for $10 \lesssim\, \rho\, \lesssim 30$ kpc, the FDM rms height grows while the CDM counterpart does not change appreciably for the same range. This happens because, in the CDM side, for the above range of $\rho$ %near central region, 
already many satellites are found with a significantly build-up value of the rms height. On the other hand, in the FDM side, there are not many satellites in that region and their rms height is building up from relatively small values.
%The estimation of SEM in this plot is calculated with varying sample number $n(\rho)$ along the distance from the center of halo system, because no satellites in some samples are identified near the center of halo system.   
%Thus 

\def\qqq6{
Note that, in the first and last panels of Figure \ref{c/a_plot},   the $\langle c/a\rangle$  functions, both for the FDM and the CDM, % at the same time, %of 
%for the two initial conditions, $(\alpha,\beta)=(0,0)$ and the corotation, 
%looks settled down 
%become 
settle down to some %approximately 
constant value around $900\, $Myr  whereas,
%those for for the other two conditions, $(\alpha,\beta)=(1,0), (1,1)$ 
in the remaining panels,
they are continuously decreasing in the late time. %, afterwards. % region.
%$\alpha=1,\beta=0$ and $\alpha=1,\beta=1$. 
For the case with $\alpha=\beta=0$, this is mainly due to the smaller kinetic energy of subhalos as %without 
any additional radial component of initial velocity is not included. %than other conditions, 
For the  corotation, this is because of the smaller total angular momentum %of system 
that was initially reduced by %from 
%the quarter  %
%the
$25$ percent 
of initial subhalos rotating in opposite direction. The cases
%cases, 
with %the initial conditions
$\alpha=\beta=0$ and  corotation also lead to a smaller galactic size than those %the systems
with the other 
two 
initial conditions.
}

%\begin{figure}[htbp]   
%\vskip-0.3cm
%\begin{center}
%\includegraphics[width=0.7\textwidth]{avr_cos_theta.png}
%\vskip-0.3cm
%\includegraphics[width=0.9\textwidth]{avr_cos_theta_outer.png}
%\caption{\footnotesize  $\A(<r)$ of satellite distribution as a function of the distance from the center %of halo 
%at $t=0.5\, \tau_c \,(\sim 1179.2$ Myr). The unit vector ${\vec{e}}_3$ of minor axis is fixed with all satellites in whole simulation box, and then $\A(<r)$ is calculated with those satellites 
%whose radial distance from the center of halo is less than $r$. 
%within the radius $r$ from the center. % of halo system.
%as explained in Section \ref{sec2.3}.
%}
%\label{avr_cos_theta_plot}
%\end{center}
%\end{figure} 

\def\qqq10{The tendency of a more flattened satellite system in the FDM halos %compared to the CDM halos 
appears also  in the angular distribution of satellites. 
In Figure \ref{avr_cos_theta_plot} we illustrate %shows
$\A(<r)$ (See (\ref{avr_cos_theta}))
%of satellites 
as a function of the distance $r$ from the center %of halo system 
at $t=0.5\, \tau_c\,(\sim 1179.2$ Myr). Note that those satellites within the radius $r$ from the center are included
as mentioned %explained 
in Section \ref{subsec:measures}. 
The difference between the FDM and the CDM model becomes %notable %rather 
%in general 
clear %generically 
and, 
for some cases, more prominent 
in the central region of halos.  
}

{
An alternative to the ratio $c/a$ is the $r$-dependent cumulative distribution function (CDF) $f_1(|\cos \theta|;r)$\footnote{Slightly abusing notation, 
we shall denote the full CDF including the whole satellites simply by $f_1(|\cos \theta|)$.} of
a satellite system where one includes only satellites within 
the radius $r$ from the center and $\cos\theta$ is given by $\hat{e}_3 \cdot \hat{r}$ with $\vec{r}$ being the position of a satellite and $\hat{e}_3$ denoting the unit eigenvector corresponding to the eigenvalue $\lambda_3$. 
We shall  use the Kolmogorov-Smirnov probability $P_{ks}$ by comparing CDF samples of the FDM/CDM
satellite systems where each CDF denoted by $f_{30}(|\cos \theta|)$ is drawn from  the angular positions of satellites in all 30 samples for each choice of initial 
%velocity 
condition\footnote{In general this quantity $P_{KS}$ represents the probability that any two sets of samples were drawn from the same cumulative probability distribution,  where the probability is determined by the so-called Kolmogorov CDF whose argument is given by $\sqrt{\frac{N_1 N_2}{N_1+N_2}} \, D_{12}$ with $N_i \,(i=1,2)$ denoting the size of the $i$-th sample and $D_{12}$ the  Kolmogorov-Smirnov statistic characterizing the maximal difference of two CDF's.}.
With the CDF's as well as
$P_{ks}$, we may deduce the main underlying factors that make the FDM halo system have a more flattened
satellite system than its CDM counterpart.
}

In Figure \ref{CDFcos_plot} we illustrate the CDF $f_{30}(|\cos \theta|)$ for the   three initial  conditions of our main interest. 
%cumulative distribution function 
%of angular positions 
%for all satellites in the full simulation box. 
%The plot on the $|\cos\theta|$ 
For each initial condition, this CDF with a subscript $30$
is made by merging all the data points in each set of  the 30 samples as explained in Section \ref{sec2.1.1}. 
The black dash-dotted line is for the CDM, the red solid  for the FDM, and the blue dashed  corresponds to the reference spherically symmetric distribution. 
%The plot on the $|\cos\theta|$ for each velocity condition is made by merging the data points of all 30 samples as described in Section \ref{sec2.1.1}. 
These plots clearly show that, for both the FDM and CDM models, the CDF's are quite distinguished from the spherically symmetric distribution.
They %These plots 
also imply that the FDM model %system
leads to more anisotropic %flattened
satellite systems than their CDM counterpart. 
%The plot in 
Figure \ref{Pks_plot} shows the $r$-dependent Kolmogorov-Smirnov probability $P_{ks}(<r)$ 
%as a function of $r$
where one includes  those satellites within the radius $r$ from the center %of halo %are included 
%for the evaluation of 
in order to build %make %compute 
%the corresponding CDF 
$f_{30}(|\cos \theta|;r)$.
 From Figure \ref{delta_plot}, one may easily guess that the Kolmogorov-Smirnov statistic $D_{12}$ may decrease as a function of $r$ especially for the region of $r$ near the centroid. However, the number of satellites increases as $r$ grows. Indeed the latter growth effect becomes  dominant, which leads to the monotonically decreasing behaviors of $P_{KS}$ as a function of $r$ as illustrated in Figure \ref{Pks_plot}.

In Table \ref{table:result}, our %flattening  
measures 
for the
%degree of 
flattening 
of satellite planes
are summarized including
the Kolmogorov-Smirnov %$P_{ks}$ 
probabilities
between the CDM and the FDM distribution.
%in the CDM and the FDM models with $P_{ks}$ probabilities 
All data are drawn from the samples at $t=1179.2$ Myr. 
%\sout{The upper nine rows are 
%for the simulations with the initial %velocity condition (\ref{initvelocity %condition}) and the bottom row is for the %simulation of the corotation test with the %initial velocity condition %(\ref{corotationvelocity condition}).
%} 
{ All of our initial conditions with $q_1=0.25, 0.3,0.35, 0.4,0.45$ are included in this table.}
These results %clearly 
show that 
the FDM and the CDM satellite system behave %indeed 
%quite 
differently from each other. %especially in their dynamical evolution.
%Our results indicate that
%the FDM and the CDM cummulative functions %CDFs, 
%including also their radial refinements, are indeed quite different from each other.
%The horizontal axis in the plot for the $P_{ks}$ represents the radius $r$ from the center of halo system, only within which the satellites are included for the CDF. 

\begin{table}[htbp]
\begin{center}
\begin{tabular}{c|cc|cc|c}
\hline\hline
 IC &  \multicolumn{2}{|c|}{c/a  (mean$\pm$std) } &\multicolumn{2}{|c|}{$\delta$ (mean$\pm$std) [kpc]}  & \multirow{2}{2em}{$P_{ks}$}\\
 \cline{1-5}
 $q_1$ &  CDM & FDM & CDM & FDM &  \\
\hline
0.25 & 0.406$\pm$0.044  & 0.223$\pm$0.073 &  9.301$\pm$0.770 &  6.872$\pm$1.987 & 1.243$\times 10^{-20}$  \\
0.30 & 0.415$\pm$0.049  & 0.222$\pm$0.063 &  9.230$\pm$0.791 &  6.737$\pm$1.789 & 1.135$\times 10^{-14}$   \\
0.35 & 0.466$\pm$0.053  & 0.236$\pm$0.067 &  9.961$\pm$0.811 &  6.832$\pm$1.875 & 1.047$\times 10^{-30}$   \\
0.40 & 0.495$\pm$0.068  & 0.267$\pm$0.108 & 10.405$\pm$1.226 &  7.548$\pm$2.676 & 2.069$\times 10^{-11}$   \\
0.45 & 0.532$\pm$0.071  & 0.295$\pm$0.120 & 10.689$\pm$1.237 &  8.068$\pm$2.735 & 4.574$\times 10^{-12}$   \\
%0.50 & 0.551$\pm$0.071  & 0.327$\pm$0.098 & 11.104$\pm$1.529 &  9.211$\pm$2.559 & 2.365$\times 10^{-11}$   \\
\hline\hline
%Co-rotation & 0.406$\pm$0.009 & 0.208$\pm$0.012 & 12.312$\pm$0.231 & 8.285$\pm$0.444 & 1.639$\times 10^{-21}$\\
%\hline
\end{tabular}
\end{center}
\vskip-0.2cm
\caption{\footnotesize 
%The comparison of the 
Here our measures for the
%degree of 
flattening of satellites
are summarized including
the Kolmogorov-Smirnov %$P_{ks}$ 
probabilities
between the CDM and the FDM distribution.
%in the CDM and the FDM models with $P_{ks}$ probabilities 
All data are drawn from the samples at $t=1179.2$ Myr. {All of our initial conditions with $q_1=0.25, 0.3,0.35, 0.4,0.45$ are included in this table.}
%The upper nine %9 
%rows are 
%for the simulations with the initial velocity condition (\ref{initvelocity condition}) and the bottom row is for the simulation 
%of 
%the %corotation test 
%with the initial velocity condition (\ref{corotationvelocity condition}). 
Note that ``IC" in this table  stands for ``initial  condition". } 
\label{table:result}
\end{table}

As mentioned in Introduction, all the satellites systems in our FDM halo models result in more flattened planes than those in the CDM counterparts with respect to all %the 
measures %adopted 
in this work.
 %\sout{The initial velocity conditions with %relatively small or no radial component %[\,$\alpha=0 $ in the velocity condition %(\ref{initvelocity condition})\,] lead to 
 %smaller differences
% in $\langle c/a \rangle$ and $\langle\delta\rangle$  (between the CDM and the FDM)
%   than the remaining 
%velocity conditions.} 
This trend may also be seen %observed
 %with the difference of 
 from the Kolmogorov-Smirnov probability $P_{ks}$ given in the Table \ref{table:result} for various initial %\sout{velocity} 
 conditions. Roughly speaking, the more similar the two distributions come to be, the larger is the corresponding Kolmogorov-Smirnov probability $P_{ks}$. % becomes.
%\sout{The main reason behind these smaller differences in  $\langle c/a \rangle$ or $\langle \delta\rangle$ for the relatively small initial radial velocities
%stems from  the smaller fractions ($0.2 \sim 0.3$) of subhalos  colliding into the central region, which is  contrasted with the cases with larger initial radial velocities (with  fractions of colliding subhalos $\sim 0.4$). 
% Notice these collisions in both cases mostly occur at early stage of the evolution processes.}   
{As $q_1$ becomes smaller, the initial collapse gets stronger, more subhalos are colliding into the central region, and the resulting
 %corresponding
 satellite system becomes more flattened.}
 Notice that these collisions 
 %\sout{in both cases} 
 mostly occur at an early stage of the evolution 
 %formation 
 processes. 
%{\cb For the both  FDM and CDM models,}  the more \sout{FDM} subhalos are subject to colliding in the central region, the more flattened the resulting
 %corresponding
 %satellite systems become. %are compared to the CDM cases. 
 %\sout{This indicates that the planar nature of the FDM halos}
 The relatively stronger flattening  in the FDM side is closely tied to the gravitational cooling effect, %of the FDM model, 
 which is especially effective around the central region of the halos.
  It should be commented that $P_{KS}$ should be correlated with the parameter $q_1$ but, in Table \ref{table:result}, it %$P_{KS}$ 
 does not increase monotonically nor does it show any clear trend with $q_1$. This is simply because $P_{KS}$ concerns about the difference between the two CDF's and the flattening trends of FDM and CDM may differ from each other.
 
We now turn to the corotation ratio {  $\eta=\frac{N^+_s}{N_s}$ where $N^\pm_s$ denotes the number of satellites involving a positive/negative $\hat\phi$-component respectively.}
{With the %corotation 
$q_1=0.35$ initial %corotation
 condition,} %(\ref{corotationvelocity condition}). In this simulation, 
one begins with { $\eta_0 = 
%{\color{red} 0.45} 
0.4922\pm 0.0197$} 
%as dictated by the $q_1=0.35$ initial  condition
both for the FDM and the CDM halo. At $t= 1179.2$ Myr,
{we find  the FDM corotation ratio $\langle\eta_{\mbox{\tiny FDM}}\rangle = 0.9240 \pm 0.0916$,
%at $t=0.5\tau_c \,(\sim 1179.2$ Myr), 
 which is bigger than the CDM value %one %ratio
$\langle\eta_{\mbox{\tiny CDM}}\rangle = 0.5888 \pm 0.0396$.} %for CDM halo 
%With this simulation, 
%The simulation with initial velocity condition (\ref{corotationvelocity condition}) gives the result that the FDM halo model have more corotating satellites systems. Specifically one finds  the FDM corotation ratio $\langle\eta_{FDM}\rangle = 0.9336 \pm 0.0098$
%at $t=0.5\tau_c \,(\sim 1179.2$ Myr), which is bigger than the CDM ratio  $\langle\eta_{CDM}\rangle = 0.8066 \pm 0.0066$. %for CDM halo 
%With this simulation, 
The resulting FDM ratio appears to be more consistent with observational data, but this requires further 
%clarifications and 
studies. % and examinations.

Finally let us comment upon %the %angular
%distributions of
satellite orbital poles, and in particular their
standard deviation \cite{Pawlowski:2014una} defined by
\begin{align}\label{delta}
\triangle_{sph} \equiv
\sqrt{
\frac{\sum^{N_s}_{k=1} [\,\arccos(|\hat{n}_{avg}\cdot\hat{n}_{(k)}|)\,]^2}{N_s}}\,,
\end{align}
where
%$N_s$ denotes the number of satellites
%and
$\hat{n}_{avg}$ is  the normalized mean vector of orbital poles given  by $\hat{n}_{avg} = \vec{n}_{avg} \,/\,\vec{n}_{avg}$ with $
\vec{n}_{avg}=(1/N_s)\sum^{N_s}_{k=1}\hat{n}_{(k)}$.
%$\hat{n}_{avg} =  1/N_s\sum^{N_s}_{k=1}\hat{n}_{(k)}\,/\,|1/N_s\sum^{N_s}_{k=1}\hat{n}_{(k)}|$
%and {\cb the sum on $l$ is over only those satellites involving a positive $\hat\phi$-component}.
%is calculated with $N_s$ satellites
%in a halo system.
%Finally we obtain the more narrow distribution of satellite orbital poles in the FDM  with smaller standard deviation of orbital poles $\langle\triangle_{sph}\rangle_{FDM} = 31.4115\pm2.383^\circ$ than the CDM counterpart $\langle\triangle_{sph}^{CDM}\rangle = 43.3645\pm0.5774^\circ$.
%for the CDM
%Note that  $\triangle_{sph}$ is calculated from the  definition
{Note that this definition is somewhat different from the usual one in  literature
where the value from  7 or 8 best aligned angular momenta of the most massive satellites
is taken.

We find a narrower distribution of  FDM satellite orbital poles with
%standard deviation of orbital poles
{$\langle\triangle^{\mbox{\tiny FDM}}_{\,sph}\rangle =
27.10\pm4.96^\circ$}
%for the FDM side
than
 the CDM side  with $\langle\triangle_{\,sph}^{\mbox{\tiny CDM}}\rangle %= 48.3645\pm0.5774^\circ$;
 = 48.91\pm2.99^\circ$;} these numbers are drawn from the samples with the
 %corotation %initial
{$q_1=0.35$} initial  condition %(\ref{corotationvelocity condition})
at $t=1179.2$ Myr.

%In summary, all the measures %clearly
%indicate that, beginning with  the same set of initial conditions, the resulting FDM 
%halos possess 
%more flattened and corotating satellite planes than %compared to
%their CDM counterparts.

%\newpage
\section{Comparison with observational data}\label{sec5}

%To fully understand the galaxy formation in the FDM model, one has to simulate the universe
%from the very early universe with computers having enormous memory and cpus.
For a thorough understanding of the galaxy formation problem, % with our FDM model, 
one has to perform a full-fledged cosmological simulation from the very early universe as was mentioned before. %previously. %already.  
Since our results in this note are based on the simplified %dark-matter only, 
toy
galaxy models %systems 
of typical size ${\cal O}(10^2)\,$kpc while adopting %with choices of  
specific initial conditions, % and dark matter only,
it is hard to compare our numerical results directly to observational data from %of
real galaxies.
Nonetheless, it is interesting to see that our toy galaxy models %systems 
in the FDM side much
resemble 
the observed %ones  %satellite systems 
%especially in their qualitative features. %
at least qualitatively
%even quantitatively to some extent 
as  discussed right below. %to some extent. %at least qualitatively.

According to observations, the satellite planes of the Milky Way typically possess an
rms height
 $\delta = 20\sim 30\,$kpc,
axis ratio
 $c/a = 0.18 \sim  0.30$, standard deviation of orbital poles
$\triangle_{sph}\simeq 25^\circ\,$\footnote{As previously mentioned, the definition of $\Delta_{sph}$ in the references is  different from ours.}
%degree,
and
inclination angle  %incl=
$\,73^\circ\sim 87^\circ$ \cite{Pawlowski:2018sys,2020MNRAS.491.3042P}
 depending on the sample.
{Similarly, the Centaurus A satellite system
has $\delta \simeq 60\,$kpc,
and
 $c/a \simeq  0.2$ \cite{2015ApJ...802L..25T}.

Our simulations %in the previous sections with
in the FDM side  typically show %have
 $c/a = 0.21 \sim  0.30$ and
$\triangle_{sph}\simeq 27^\circ\negthinspace$ (with the
{$q_1=0.35$} initial  condition), %degree,
which well agree %well
with the above observed values for the Milky Way.
%are indeed quite similar to the observed values.
Our simulations in the CDM side, on the contrary,   exhibit typically
 $c/a = 0.41 \sim  0.53$ and $\triangle_{sph}\simeq 49^\circ$ (with the
{$q_1=0.35$} initial  condition),
 which are larger than the observed ones. %data.
  These $c/a$ values for the CDM are  consistent
 with those of the planes of satellite galaxies in the EAGLE simulation
 with the CDM \cite{Shao:2019nuc}.
 Thus one may say that the FDM model works better at reproducing
 the observed satellite plane structures than the CDM counterpart.
 }

On the other hand, the rms heights for the both DM models are around $\delta\simeq 10\,$kpc, which %is certainly 
appears smaller than the observed. %ones. %values.
This discrepancy may be attributed %due 
to the relatively smaller size of the simulated centroid (carrying a typical
mass $M_{tot}\simeq 5\times 10^{10}\, M_\odot$)
%in comparison with %
than that of %compared to 
the observed  central galaxies
possessing a heavier typical
mass $M_{tot}={\cal O}(10^{12})\, M_\odot$. 
Assuming a similar average DM density
for all galaxies, the spatial size of the heavier galaxies
may become 
%$(10^{12}/5\times 10^{10})^{1/3}\sim 2.7$
$2.7$ [\,$\sim (10^{12}/5\times 10^{10})^{1/3}$\,]
times larger than the simulated value.
Therefore, we expect $\delta\simeq 27\,$kpc for the heavier galaxies,
which is indeed comparable to the observed.
To confirm this we need a simulation with a larger box,
which is beyond the scope of this paper.
% values.
%It should also be noted that our simulations do not have enough resolutions to identify
%the inclination angle of the 
%central regions.

We interpret our numerical results in the following way.
{Energy loss with a conserved angular momentum during a collapse usually leads to a formation of a highly flattened astronomical structure. 
  In both dark matter models, during the collapse satellites gravitationally exchange their energy and momentum with others
  and experience a kind of an averaging process. 
Those satellites getting a relatively large angular momentum during the process may escape the central region and easily survive,
while  satellites with a small angular momentum tend to fall into the central region. 

In the FDM side satellites which  get
small angular momentum and fall
into the central region  can be absorbed into
%the halo of 
the centroid 
%(central galaxy) 
by losing their energy  via the gravitational cooling  effect.
Apparently, in our simulations a large portion of the falling satellites in the FDM side
are tidally disrupted and lose their cores and  disperse completely. 
As a result surviving satellites  in the FDM side have a tendency 
to form corotating planar structures orthogonal to the total angular momentum vector.
Note that this effect is clearly seen in Figure \ref{5x4DFmap}.

On the other hand,  a large portion of
similar falling satellites in the CDM side
may  survive the crossing through the centroid %with 
 in the same situation, mainly due to the particle nature of the CDM model.
 Thus, we expect satellites in the CDM model without an efficient energy loss mechanism
 redistribute themselves more spherically than in the FDM model as seen
 in Figure \ref{mgsDFmap}.
%On the other hand, }
%Satellites moving mainly in the $z$-direction usually
%have a loin our coordinate system, hence there are higher chances
%for them to be absorbed into the centroid. In the FDM model, this absorption process  is
%more efficient than that of the CDM, % model,  
%which is mainly due to the gravitational cooling 
%%and tidal 
%effect in the FDM side. }

In our scenario we also expect that
the width of a satellite plane in the FDM side
could be roughly related to the size of the centroid, because
the spatial size of the centroid provides a typical  length scale for the satellite absorption process near
the centroid which is responsible for the energy loss and the formation of the satellite planes
as previously mentioned.

Another interesting observation is that, even with the same initial conditions,
there are less FDM satellites surviving 
compared to the CDM side due to the same effects making satellite planes.
This hints yet another route in the FDM model to solve the missing satellite problem of the CDM model,
because  the typical solution  in the FDM model
to the problem is usually attributed to the suppression of the initial matter spectrum \cite{Fuzzy} compared to the CDM model. 
}

For more conclusive results we need more realistic simulations for larger
galaxies, which is beyond the scope of this work.
Since our simulations are DM-only simulations, they do not provide information of
 inclination angles of the satellite planes with
respect to the galactic disks. 

\section{Conclusions}\label{sec6}

%We numerically showed a tendency that 
%satellite galaxy systems in the FDM model is more flattened
%   than in the CDM model due to dissipation from gravitational cooling effect near a central part of galaxy.
In this note, we have numerically shown 
that galactic  satellite  systems in the FDM model are more flattened and corotating than their CDM counterpart. This is basically 
due 
%an extra dissipation of 
the gravitational cooling effect   in the FDM side especially near the central part of galaxies. 
{This energy loss mechanism is a unique feature of the FDM  which
is absent in other alternatives of the CDM such as warm dark matter.}
Our toy %model 
galaxies in the FDM model seem to reproduce
the observational features such as axis ratios of the satellite planes. 
%close %similar 
%to the observed data.

Our work implies that
the FDM model could be a way to solve the problem of the %plane of 
satellite-galaxy planes as well as other small scale tensions. %problem,
Thus the galactic satellite planes could serve as a good test bed for dark matter model discrimination.
For more conclusive results,
cosmological %scale 
simulations with baryons starting from more realistic initial conditions
are certainly required, which we leave for our future studies.

%Our work also implies another solution to missing satellite problem of the CDM model.
%The solution of the missing satellite problem with the FDM model is often
%attributed to the suppression of initial power spectrum at galactic scales
% compared to the CDM model.
%However, here we have demonstrated that, even with the same initial seed spectrum,
%the gravitational cooling of the FDM model  near the central part of galaxy 
%suppresses the survival of satellite galaxies as contrary to its CDM counterpart.

\subsection*{Acknowledgement}
We would like to thank   the referee for very  constructive suggestions %comments 
%that helped to
%improve the paper. 
and Andreas Gustavsson for careful reading of the manuscript.
Sangnam Park was supported in part 
%NRF Grant 2020R1A2B5B01001473, and
by  Basic Science Research Program
through National Research Foundation funded by the Ministry of Education (2018R1A6A1A06024977).
Dongsu Bak was supported in part by NRF Grant 2020R1A2B5B01001473, and by  Basic Science Research Program
through National Research Foundation funded by the Ministry of Education (2018R1A6A1A06024977).
Jae-Weon Lee was supported by NRF-2020R1F1A1061160.
Inkyu Park was supported by the 2021 Research Fund of the University of Seoul.
This work was also supported by the UBAI computing resources 
%of Urban Big data and AI Institute (UBAI)
at the University of Seoul.

\end{document}